\documentclass[floats,nofootinbib,preprint,superscriptaddress,tightenlines]{revtex4}
\pdfoutput=1

\usepackage{amsfonts,amsmath,amssymb}
\usepackage{bm}
\usepackage[colorlinks,linkcolor=blue,anchorcolor=black,citecolor=blue]{hyperref}
\usepackage{graphicx}
\usepackage{slashed}
\usepackage{xcolor}

\begin{document}
\baselineskip=17pt \parskip=5pt

\preprint{NCTS-PH/2005}
\hspace*{\fill}

\title{Evading the Grossman-Nir bound with \boldmath$\Delta I=3/2$ new physics}

\author{Xiao-Gang He}\email{hexg@phys.ntu.edu.tw}
\affiliation{Department of Physics, National Taiwan University,\\
No.\,\,1, Sec.\,\,4, Roosevelt Rd., Taipei 106, Taiwan}
\affiliation{Physics Division, National Center for Theoretical Sciences,\\
No.\,\,101, Sec.\,\,2, Kuang Fu Rd., Hsinchu 300, Taiwan}

\author{Xiao-Dong Ma}\email{maxid@phys.ntu.edu.tw}
\affiliation{Department of Physics, National Taiwan University,\\
No.\,\,1, Sec.\,\,4, Roosevelt Rd., Taipei 106, Taiwan}

\author{Jusak Tandean}\email{jtandean@phys.ntu.edu.tw}
\affiliation{Department of Physics, National Taiwan University,\\
No.\,\,1, Sec.\,\,4, Roosevelt Rd., Taipei 106, Taiwan}
\affiliation{Physics Division, National Center for Theoretical Sciences,\\
No.\,\,101, Sec.\,\,2, Kuang Fu Rd., Hsinchu 300, Taiwan}

\author{German Valencia}\email{german.valencia@monash.edu}
\affiliation{School of Physics and Astronomy, Monash University,\\
Wellington Road, Clayton, VIC-3800, Australia
\bigskip}


\begin{abstract}

Rare kaon decays with missing energy, $K\to\pi$+$E_{\rm miss}$, have received considerable attention because their rates can be calculated quite precisely within the standard model (SM), where the missing energy is carried away by an undetected neutrino-antineutrino pair.
Beyond the SM, clean theoretical predictions can also be made regarding these processes.
One such prediction is the so-called Grossman-Nir (GN) bound, which states that the branching fractions of the $K_L$ and $K^+$ modes must satisfy the relation $\mathcal{B}(K_L\to\pi^0$+$E_{\rm miss})\lesssim4.3\,\mathcal{B}(K^+\to\pi^+$+$E_{\rm miss})$ and applies within and beyond the SM, as long as the hadronic transitions change isospin by $\Delta I=1/2$.
In this paper we extend the study of these modes to include new-physics scenarios where the missing energy is due to unobserved lepton-number-violating neutrino pairs, invisible light new scalars, or pairs of such scalars.
The new interactions are assumed to arise above the electroweak scale and described by an effective field theory.
We explore the possibility of violating the GN bound through $\Delta I=3/2$ contributions to the $K\to\pi$ transitions within these scenarios and find that large violations are only possible in the case where the missing energy is due to an invisible light new scalar.

\end{abstract}

\maketitle

{\small\hypersetup{linkcolor=black}\tableofcontents}

\newpage

\section{Introduction\label{intro}}

The rare kaon decays $K_L\to \pi^0\nu\bar\nu$ and $K^+\to \pi^+\nu\bar\nu$ are known as the ``golden modes'' of kaon physics because they can be predicted quite precisely within the standard model (SM) and are potentially sensitive to the presence of new physics (NP) beyond it \cite{Littenberg:1989ix,Hagelin:1989wt}. Consequently they have received a great deal of theoretical and experimental attention. The SM predictions for their branching fractions have been known for many years \cite{Mescia:2007kn,Inami:1980fz,Buchalla:1998ba,Brod:2010hi,Buras:2006gb,Brod:2008ss,Isidori:2005xm,Falk:2000nm}, and their current values are~\cite{Charles:2004jd,ckmfitter,Tanabashi:2018oca} ${\mathcal B}(K_L\to\pi^0\nu\bar\nu)_{\rm SM} = (3.0\pm0.2)\times 10^{-11}$ and ${\mathcal B}(K^+\to\pi^+\nu\bar\nu)_{\rm SM} = (8.5\pm 0.5)\times 10^{-11}$, which suffer mostly from parametric uncertainties. These numbers, and their counterparts in many NP scenarios, respect a particularly interesting prediction, the so-called Grossman-Nir bound~\cite{Grossman:1997sk}, which states that $\mathcal{B}(K_L\to\pi^0\nu\bar\nu)\lesssim 4.3\,\mathcal{B}(K^+\to\pi^+\nu\bar\nu)$. A key assumption behind this statement is that the $K\to \pi$ transition proceeds from an interaction which changes isospin by $\Delta I = 1/2$.

On the experimental side, the KOTO Collaboration at J-PARC in 2018 set an upper limit on the neutral mode based on the data collected in 2015~\cite{Ahn:2018mvc}: ${\mathcal B}(K_L\to\pi^0\nu\bar\nu)_{\rm KOTO15}<3.0\times10^{-9}$ at 90\% confidence level (CL). For the charged mode, the combined BNL E787/E949 experiments had earlier yielded ${\mathcal B}(K^+\to\pi^+\nu\bar\nu)_{\rm E949}=\big(1.73_{-1.05}^{+1.15}\big) \times 10^{-10}$~\cite{Artamonov:2008qb,Artamonov:2009sz}.
Last year the NA62 Collaboration~\cite{CortinaGil:2018fkc} at CERN reported the preliminary limit ${\mathcal B}(K^+\to\pi^+\nu\bar\nu)_{\rm NA62} < 1.85\times 10^{-10}$ at 90\% CL~\cite{na62}.
All of these results are in agreement with the SM but leave open a window for~NP. This possibility is very intriguing, especially in light of the recent preliminary observation by KOTO of three candidate events in the $K_L\to \pi^0\nu\bar\nu$ signal region~\cite{koto}, with a single event sensitivity of $6.9\times 10^{-10}$ having been achieved.
If interpreted as signal, they imply a decay rate about two orders of magnitude higher than the SM prediction and in conflict with the experimentally established GN bound ${\mathcal B}(K_L\to\pi^0\nu\bar\nu)_{\rm GN}<4.3\,{\mathcal B}(K^+\to\pi^+\nu\bar\nu)_{\rm NA62}=8.0\times10^{-10}$ at 90\% CL.

It is interesting that there are NP scenarios which can overcome this requisite. For instance, as was first pointed out in ref.\,\cite{Fuyuto:2014cya}, the branching fraction of $K_L\to\pi^0\textsl{\texttt X}$, with \textsl{\texttt X} being an invisible particle with mass $m_{\textsl{\texttt X}}$ chosen to be around the pion mass, can exceed the aforementioned cap of $8.0\times10^{-10}$ because quests for the charged channel $K^+\to\pi^+\textsl{\texttt X}$ do not cover the $m_{\textsl{\texttt X}}\sim m_\pi$ region to avoid the sizable $K^+\to\pi^+\pi^0$ background~\cite{Artamonov:2009sz,CortinaGil:2018fkc}. This kinematic loophole has been exploited in recent attempts~\cite{Kitahara:2019lws,Egana-Ugrinovic:2019wzj,Dev:2019hho,Liu:2020qgx,Liao:2020boe,Jho:2020jsa,Cline:2020mdt} to account for KOTO's anomalous events.
As another example, imposing the condition $m_{K^+}-m_{\pi^+}<m_{\cal X}<m_{K^0}-m_{\pi^0}$ renders the $K^+\to\pi^+\cal X$ channel closed (here $\cal X$ can be more than one invisible particle), whereas $K_L\to\pi^0\cal X$ with a big rate can still happen \cite{Fabbrichesi:2019bmo}.
For other $m_{\cal X}$ ranges, the stringent restriction on $K^+\to\pi^+\cal X$ can be evaded or weakened if $\cal X$ is a particle with an average decay length bigger than the KOTO detector size but less than its E949 and NA62 counterparts~\cite{Kitahara:2019lws,Egana-Ugrinovic:2019wzj,Dev:2019hho,Liu:2020qgx,Liao:2020boe}.
Most of these cases, while still fulfilling the GN relation, only appear to contradict it by enhancing ${\mathcal B}(K_L\to\pi^0\cal X)$ substantially above~${\mathcal B}(K_L\to\pi^0\nu\bar\nu)_{\rm GN}^{\rm max}$.
In a framework of effective field theory with only SM fields, the introduction of (predominantly) $\Delta I = 1/2$ interactions which change lepton flavor/number or possess new sources of $CP$ violation also would not bring about a disruption of the GN bound~\cite{Li:2019fhz,He:2020jzn}.
On the other hand, it has been proposed that in the presence of additional light particles mediating these decays the GN bound could be violated~\cite{Ziegler:2020ize}.

In terms of the branching-fraction ratio \,$r_{\cal B} = {\cal B}(K_L\to\pi^0\nu\bar\nu)/{\cal B}(K^+\to\pi^+\nu\bar\nu)$,\, the GN bound has a theoretical model-independent maximum of \,$r_{\cal B}^{\rm GN}=4.3$,\, which differs much from the central value \,$r_{\cal B}^{\rm SM}= 0.36$\, of the SM prediction.
A key ingredient in the derivation of the GN inequality is the assumption that the $K\to\pi$ transitions are mediated by a two-quark $s\leftrightarrow d$ operator, which necessarily carries isospin $1/2$ and leads to a ratio of amplitudes for the neutral and charged modes given by \,${\cal A}^{\Delta I = 1/2}_{K^0\to \pi^0}/{\cal A}^{\Delta I = 1/2}_{K^+\to \pi^+} = -1/\sqrt{2}$.\,
A true violation of the GN relation requires the $K \to \pi$ transitions to occur via a $\Delta I=3/2$ interaction as well, and this possibility has recently been investigated in refs.~\cite{He:2018uey,He:2020jzn}.
A pure \,$\Delta I = 3/2$\, operator would result in an amplitude ratio ${\cal A}^{\Delta I =3/2}_{K^0\to \pi^0}/{\cal A}^{\Delta I = 3/2}_{K^+\to \pi^+}=\sqrt2$\, and thus translate into \,$r_{\cal B}^{\Delta I=3/2} \lesssim 17$ \cite{He:2020jzn}.

In this paper we present a systematic study on how to overcome the GN bound with quark-level
operators in the context of effective field theory (EFT). Given that the measurements of interest
look for \,$K\to\pi\cal X$\, with $\cal X$ standing for one or more particles carrying away missing
energy\,\,($E_{\rm miss}$), we will consider for $\cal X$ several different possibilities:
a neutrino-antineutrino pair\,\,($\nu\bar\nu$), a\,\,pair of neutrinos ($\nu\nu$) or antineutrinos
($\bar\nu\bar\nu$), an invisible light new scalar boson ($S$), and a pair of these scalars ($SS$).
Since an operator directly giving rise to \,$\Delta I=3/2$\, \,$K\to\pi\cal X$\, transitions has to
contain at least four light-quark fields, the minimal mass dimension of such an operator is seven,
eight, nine, and ten for \,${\cal X}=S,\,SS$, $\nu\nu{\rm\;or\;}\bar\nu\bar\nu$, and $\nu\bar\nu$,
respectively.
Being unobserved in the experiments, the neutrinos may have different flavors and can also be
replaced by new invisible light fermions.
We suppose that the new particles are invisible because they are sufficiently long-lived to escape
detection, decay invisibly, or are stable.
We will look at a complete set of operators with the lowest dimension necessary for each of these
cases, assuming that the interactions of the new light particles with the quarks can be described
by an EFT approach valid above the electroweak-symmetry breaking scale.

Before embarking in a detailed study, it is instructive to  present a simple dimensional-analysis estimate for the scale $\Lambda_{\textsc{np}}$ of NP that is needed in order to produce a rate of $K\to\pi\cal X$ that is comparable in size to the SM rate. To this end, it is useful to recall the effective Hamiltonian responsible for $K\to\pi\nu\bar\nu$ in the SM,
\begin{align}
{\cal H}_{\rm SM} & \,=\, \frac{G_{\textsc f}}{\sqrt{2}}\frac{g^2V^*_{ts}V_{td}^{} X(x_t)}{16\pi^2} \bar{s}\gamma_\mu(1-\gamma_5^{})d \sum_\ell \bar{\nu}_\ell\gamma^\mu(1-\gamma_5^{})\nu_\ell \,+\, {\rm H.c.}\,, \label{smH}
\end{align}
in the conventional notation~\cite{Tanabashi:2018oca}, where $X(x_t)\simeq1.5$ from top-quark loops. In contrast, a generic NP operator which induces a $\Delta I=3/2$ transition in the process $K \to \pi\cal X$ can be written as
\begin{align}
{\cal L}_{\rm NP} & \,=\, \frac{1}{\Lambda_{\textsc{np}}^{2+{\textsc n}_{\cal X}}}~{\cal C}~\bar{s}\Gamma_1u \bar{u}\Gamma_2 d~\cal X \,+\, {\rm H.c.} \,,
\end{align}
where ${\textsc n}_{\cal X}$ tells the mass dimension of $\cal X$ (so ${\textsc n}_{\nu\nu}=3$, ${\textsc n}_S=1$, ${\textsc n}_{SS}=2$, etc.), the coefficient $\cal C$ is a constant, and $\Gamma_{1,2}$ represent gamma matrices.
At the amplitude level, the ${\cal L}_{\rm NP}$ contribution to $K\to\pi\cal X$ relative to the SM top-quark contribution is then
\begin{align}
\frac{{\cal A}_{\rm NP}}{{\cal A}_{\rm SM}} & \,\sim\, 3.8\times 10^5~ {\cal C}~ \left(\frac{v}{\Lambda_{\textsc{np}}}\right)^2 \left(\frac{m_K}{\Lambda_{\textsc{np}}}\right)^{{\textsc n}_{\cal X}} , \label{scale}
\end{align}
where the large numerical factor reflects the one-loop and CKM-angle suppression of the SM coefficient, $v=2^{-1/4}G_{\textsc f}^{-1/2}=246$\,\,GeV indicates the electroweak scale, and the relevant hadronic scale is taken to be the kaon mass, $m_K$. If the NP is defined to enter ${\cal L}_{\rm NP}$ with $\cal C$ of order one (so any loop or mixing-angle suppression factor is absorbed into $\Lambda_{\textsc{np}}$), the result in eq.\,(\ref{scale}) implies that $\Lambda_{\textsc{np}}\sim 2.2$\,\,TeV, 275\,\,GeV, 78\,\,GeV for ${\textsc n}_{\cal X}=1,2,3$, respectively, correspond to NP effects at the same level as the SM contribution.

The EFT that we employ in this paper only makes sense for $\Lambda_{\textsc{np}}>v$. Otherwise, the organization of effective operators in terms of their dimensionality breaks down. This suggests that scenarios with ${\cal X}=S$ could amplify the rate of $K_L\to \pi^0\cal X$ to the level implied by the three KOTO events; scenarios with ${\cal X}=SS$ may increase the corresponding rate compared to its SM counterpart by factors of 2; and that scenarios in which ${\cal X}=\nu\nu{\rm\;or\;}\bar\nu\bar\nu{\rm \;or\;}\nu\bar\nu$ can only modify the $K\to\pi$+$E_{\rm miss}$ rates marginally.

We will explore all these possibilities in detail to quantify how the underlying NP interactions influence these kaon decays.
To handle a low-energy process involving hadrons, it is necessary to hadronize the quark-level operators at the mass scale where it takes place.
For our investigation, this entails computing the effects of QCD renormalization group (RG) running from the electroweak (EW) scale to the chiral-symmetry breaking scale and subsequently matching the resulting operators to a low-energy chiral Lagrangian suitable to describe the $K\to\pi$+$E_{\rm miss}$ transitions.
In all the cases, we present numerical results illustrating the range of values which the ratio $r_{\cal B}$ can take when the NP scale is under a couple of TeV.

The arrangement of the rest of the paper is as follows. In section \ref{k2pnn} we study how the GN bound can be violated through $K\to\pi\nu\nu,\pi\bar\nu\bar\nu$ caused by $\Delta I=3/2$ interactions in the EFT framework where the operators respect the SM gauge symmetries (SMEFT). In section \ref{k2ps} we extend the SMEFT with the addition of a SM-gauge-singlet scalar $S$ and carry out a similar analysis with $K\to\pi S,\pi SS$.
In section \ref{concl}, we draw our conclusions. We relegate some extra details to appendixes, including the RG running of the four-quark parts of the operators pertaining to $K\to\pi S(S)$ and their low-energy chiral bosonization.

\section{GN-bound violation via EFT operators for \boldmath$K\to\pi2\nu$\label{k2pnn}}

\subsection{\bf EFT operators for \boldmath$K\to\pi2\nu$\label{k2p2n}}

In this section we restrict the fields of the EFT to only those in the minimal SM.
Accordingly, in the kaon decays of concern the missing energy is carried away by a pair of SM neutrinos~($2\nu$). It may have no or nonzero lepton number depending on whether the underlying interaction is lepton-number conserving or violating, respectively.
If the small contributions to $K\to\pi2\nu$ from long-distance physics~\cite{Li:2019fhz,Hagelin:1989wt,Rein:1989tr,Buchalla:1998ux,Lu:1994ww} are neglected,\footnote{We also ignore isospin-breaking effects due to the $u$- and $d$-quark mass difference and electroweak radiative corrections.\medskip} the only possible way to break the GN bound significantly in this case is through the inclusion of $\Delta I=3/2$ operators having nonleptonic parts with at least four quark fields~\cite{He:2018uey,He:2020jzn}.
It follows that the lowest dimension of quark-neutrino operators with $\Delta I=3/2$ components is nine.

In the SMEFT treatment the effective operators constructed must be singlets under the SM gauge group SU(3)$_C\times$SU(2)$_L\times$U(1)$_Y$, while in the low-energy effective field theory (LEFT) the operators are to be singlets only under the strong and electromagnetic gauge group SU(3)$_C\times$U(1)$_{\rm em}$.
This means that there are in general more requirements on the SMEFT operators than on the LEFT ones, which makes the number of quark operators relatively less in the former case. Since furthermore there is still no discovery of new particles beyond the SM, we will rely on the SMEFT to perform our examination.

The fundamental fields (with their SM gauge group assignments) available to construct the pertinent operators are the U(1)$_Y$ gauge field \,$B\,(1,1,0)$,\, the SU(2)$_L$ gauge field \,$W(1,3,0)$,\, the SU(3)$_C$ gluon field \,$G\,(8,1,0)$,\, the Higgs doublet \,$H\,(1,2,1/2)$,\, the left-handed quark doublet \,$Q\,(3,2,1/6)$,\, the right-handed quark fields \,$u\,(3,1,2/3)$\, and \,$d\,(3,1,-1/3)$,\, the left-handed lepton doublet \,$L\,(1,2,-1/2)$,\, and the right-handed charged lepton field \,$e\,(1,1,-1)$.\, All the quarks and leptons come in three families. In the SMEFT approach, operators with four quarks and two neutrinos are necessarily of dimension nine (dim-9) or higher.

If lepton number is conserved in the decays of concern,\footnote{In this and the next paragraph, we suppress the family labels of the SM fermion fields. Generally these processes may change quark and/or lepton flavors.} the responsible operator necessarily involves a pair of $L$ and $\bar L$, which supplies the $\nu\bar\nu$ pair in $K\to\pi\nu\bar\nu$.  
One can always utilize some appropriate Fierz relations to arrange the lepton fields such that they appear in the operators in the form $\bar L\gamma_\mu L$ or $\bar L{\cal D}_\mu\gamma_\rho L$, with ${\cal D}_\mu$ being a covariant derivative. To join $\bar L\gamma_\mu L$ with four quark fields to form an operator that is a singlet under the SM gauge group, the quark portion needs to have a Lorentz index to contract with the one in the lepton current. The possible lowest-dimension quark parts are $(\bar Q \gamma^\mu Q,\bar{\textsl{\textsf q}}\gamma^\mu\textsl{\textsf q})(\bar Q\textsl{\textsf q},\bar{\textsl{\textsf q}}Q)$, where $\textsl{\textsf q}$ may be $u$ or $d$ as appropriate, but they have odd numbers of $Q$ and hence are not SU(2)$_L$ singlets yet. These quark combinations can be made singlets by incorporating the Higgs field $H$, and so the full quark-lepton operators have dimension ten (dim-10). Additionally, one can insert ${\cal D}^\mu$ in $(\bar Q\textsl{\textsf q},\bar{\textsl{\textsf q}}Q)(\bar Q\textsl{\textsf q},\bar{\textsl{\textsf q}}Q)$ to form singlets, and the resulting operators are also of dim-10.
If the lepton bilinear is $\bar L{\cal D}_\mu\gamma_\rho L$ instead, the lowest-dimensional possibilities of the four-quark portion are $\bar{\textsl{\textsf q}}\gamma^\mu\textsl{\textsf q}\bar{\textsl{\textsf q}}\gamma^\rho\textsl{\textsf q}$, $g^{\mu\rho}\bar Q\textsl{\textsf q}\bar{\textsl{\textsf q}}Q$, $\bar Q\sigma^{\mu\rho}\textsl{\textsf q}\bar{\textsl{\textsf q}}Q$, $\bar Q\textsl{\textsf q}\bar{\textsl{\textsf q}}\sigma^{\mu\rho}Q$, and $\bar Q\gamma^\mu Q\bar Q\gamma^\rho Q$. Again the singlet quark-lepton operators constructed are of dim-10.

If lepton-number violation is allowed, the situation is different, as we have $K\to\pi\nu\nu,\pi\bar\nu\bar\nu$. The lepton bilinear of the lowest dimension can be organized in the form $\bar LL^{\textsc c}$ $\big(\bar L\sigma_{\mu\rho}L^{\textsc c}\big)$ or its Hermitian conjugate, the superscript {\textsc c} signifying charge conjugation. One can attach the bilinear to $\bar Q\textsl{\textsf q}\bar{\textsl{\textsf q}}Q$ or $\bar{\textsl{\textsf q}}Q\bar{\textsl{\textsf q}}Q$ $\big(\bar Q\sigma^{\mu\rho}\textsl{\textsf q}\bar{\textsl{\textsf q}}Q$, $\bar Q\textsl{\textsf q}\bar{\textsl{\textsf q}}\sigma^{\mu\rho}Q$, or $\bar{\textsl{\textsf q}}\sigma^{\mu\rho}Q\bar{\textsl{\textsf q}}Q\big)$ to form a SM-gauge singlet. Thus, the resulting operators have dim-9, which is less than that of the lepton-number-conserving ones mentioned in the previous paragraph. Hereafter in this section, we concentrate on the dim-9 operators and later briefly comment on the dim-10 case.

It has been shown that all dim-9 SMEFT operators do not preserve lepton and/or baryon numbers~\cite{Kobach:2016ami}.
We will not be interested in the baryon-number violating ones, as our aim is to study how dim-9 operators give rise to $K\to\pi\nu\nu,\pi\bar\nu\bar\nu$ and can violate the GN bound.
In the following we enumerate all of those containing one strange quark, $s$, and two neutrinos or antineutrinos. Upon imposing the SM gauge symmetries and applying Fierz transformations, we find the independent operators that can induce $K\to \pi \bar\nu\bar\nu$ to be\footnote{The factorization of the quark and lepton components is guaranteed by the application of Fierz transformations.}
\begin{align}\nonumber \hphantom{O_n}
{\cal O}_1^{opxy,\alpha\beta} &= \epsilon_{ij}\delta_{kl}( \overline{Q_o^k}\gamma_\mu Q_p^j)( \overline{d_x}\gamma^\mu u_y ) \overline{L_{\{\alpha}^{{\textsc c} i}}L^l_{\beta\}} \,,
\\\nonumber
\tilde{\cal O}_1^{opxy,\alpha\beta} &= \epsilon_{ij}\delta_{kl}( \overline{Q_o^k}\gamma_\mu Q_p^j][ \overline{d_x}\gamma^\mu u_y ) \overline{L_{\{\alpha}^{{\textsc c}i}}L^l_{\beta\}} \,,
\\\nonumber
{\cal O}_2^{opxy,\alpha\beta} &= \epsilon_{ij}\delta_{kl}( \overline{Q_o^k}\gamma_\mu Q_p^j)( \overline{d_x}\gamma_\rho u_y ) \overline{L_{[\alpha}^{{\textsc c}i}}\sigma^{\mu\rho}L^l_{\beta]} \,,
\\\nonumber
\tilde{\cal O}_2^{opxy,\alpha\beta} &= \epsilon_{ij}\delta_{kl}( \overline{Q_o^k}\gamma_\mu Q_p^j][\overline{d_x}\gamma_\rho u_y ) \overline{L_{[\alpha}^{{\textsc c}i}}\sigma^{\mu\rho}L^l_{\beta]} \,,
\\\nonumber
{\cal O}_3^{opxy,\alpha\beta} &= \epsilon_{ik}\epsilon_{jl} ( \overline{d_o}Q_p^i)( \overline{d_x}Q_y^j )\overline{L_{\{\alpha}^{{\textsc c}k}}L_{\beta\}}^l \,,
\\\nonumber
\tilde{\cal O}_3^{opxy,\alpha\beta} &= \epsilon_{ik}\epsilon_{jl} (\overline{d_o}Q_p^i][\overline{d_x}Q_y^j) \overline{L_{\{\alpha}^{{\textsc c}k}}L_{\beta\}}^l \,,
\\\nonumber
{\cal O}_4^{opxy,\alpha\beta} &= \epsilon_{ik}\epsilon_{jl} ( \overline{d_o}Q_p^i)( \overline{d_x}\sigma_{\mu\omega}Q_y^j ) \overline{L_{[\alpha}^{{\textsc c}k}}\sigma^{\mu\omega} L_{\beta]}^l \,,
\\\label{opebasis1}
\tilde{\cal O}_4^{opxy,\alpha\beta} &= \epsilon_{ik}\epsilon_{jl} ( \overline{d_o}\sigma_{\mu\rho}Q_p^i)( \overline{d_x}\sigma_{\omega}^{~\rho}Q_y^j ) \overline{L_{[\alpha}^{{\textsc c}k}}\sigma^{\mu\omega} L_{\beta]}^l \,,
\end{align}
where $o,p,x,y$ ($\alpha,\beta$) denote quark (lepton) family indices, summation over the SU(2)$_L$ indices $i,j,k,l=1,2$ is implicit, and the leptonic scalar (tensor) currents have been arranged to be symmetric (antisymmetric) in their family indices with the convention $A_{\{\alpha}B_{\beta\}}=(A_\alpha B_\beta+A_\beta B_\alpha)/2$ and $A_{[\alpha}B_{\beta]}=(A_\alpha B_\beta-A_\beta B_\alpha)/2$. The two brackets ( , ) and [ , ] in the quark bilinears distinguish the two different ways of color contraction in the products of four quark fields to form color invariants: $\bar q_1^{\textsf m}q_2^{\textsf m}\bar q_3^{\textsf n}q_4^{\textsf n}=(\bar q_1q_2)(\bar q_3q_4)$ and $\bar q_1^{\textsf m}q_2^{\textsf n}\bar q_3^{\textsf n}q_4^{\textsf m}=(\bar q_1q_2][\bar q_3q_4)$, with the color labels ${\textsf m},{\textsf n}=1,2,3$ being summed over. Each of the operators is accompanied by an unknown Wilson coefficient $C_i$, so that $C_1^{opxy,\alpha\beta}$ belongs to ${\cal O}_1^{opxy,\alpha\beta}$, etc.
The Hermitian conjugates of ${\cal O}_{1,2,3,4}^{opxy,\alpha\beta}$ and $\tilde{\cal O}_{1,2,3,4}^{opxy,\alpha\beta}$ contribute to $K\to\pi\nu\nu$.

As already stated, we assume that the SMEFT operators in eq.\,(\ref{opebasis1}) arise from NP above the EW scale.
Consequently, to address their potential impact on $K\to\pi\nu\nu,\pi\bar\nu\bar\nu$, we will first take into account the QCD effects causing the coefficients to evolve from the NP scale down to the hadronic scale, which we select to be the conventional chiral-symmetry breaking scale $\Lambda_\chi=4\pi F_\pi\simeq1.2\;$GeV with $F_\pi$ being the pion decay constant.
Subsequently, we will rely on chiral perturbation theory~\cite{Gasser:1983yg, Gasser:1984gg, Kambor:1989tz}, in conjunction with spurion techniques~\cite{Graesser:2016bpz,Cirigliano:2017ymo,Liao:2019gex}, to derive the meson-neutrino operators which contribute to the amplitudes for the kaon decays.
Based on the chiral power-counting arguments in this procedure, the operators in eq.\,(\ref{opebasis1}) with the leptonic scalar density are of momentum order $p^0$, whereas those with the leptonic tensor current are of order $p^2$.
This implies that the latter operators yield contributions to the amplitudes which are suppressed relative to those of the former by the factor $p_K^{}p_\pi^{}/\Lambda_\chi^2\sim0.05$.
Therefore, upon singling out the operators in eq.\,(\ref{opebasis1}) pertaining to $K\to\pi\nu\nu,\pi\bar\nu\bar\nu$ and neglecting those with the leptonic tensor current, in what follows we can focus on
\begin{align}\nonumber
{\texttt O}_1^{usdu}&=( \overline{u_L}\gamma_\mu s_L)(\overline{d_R}\gamma^\mu u_R)J\,, &
\tilde{\texttt O}{}_1^{usdu}&=( \overline{u_L}\gamma_\mu s_L][\overline{d_R}\gamma^\mu u_R)J\,,
\\\nonumber
{\texttt O}_1^{udsu}&=( \overline{u_L}\gamma_\mu d_L)( \overline{s_R}\gamma^\mu u_R)J\,, &
\tilde{\texttt O}{}_1^{udsu}&=( \overline{u_L}\gamma_\mu d_L][\overline{s_R}\gamma^\mu u_R)J\,,
\\\nonumber
{\texttt O}_3^{ddds}&=(\overline{d_R}d_L)(\overline{d_R}s_L)J\,, &
\tilde{\texttt O}{}_3^{ddds}&=(\overline{d_R}d_L][\overline{d_R}s_L)J\,,
\\
{\texttt O}_3^{ddsd}&=(\overline{d_R}d_L)(\overline{s_R}d_L)J\,, &
\tilde{\texttt O}{}_3^{ddsd}&=(\overline{d_R}d_L][\overline{s_R}d_L)J\,,   \label{opebasis2}
\end{align}
where $f_{L,R}^{}=(1\mp\gamma_5)f/2$ and the neutrino part is expressed as $J=( \overline{\nu_\alpha^{\textsc c}}\nu_{\beta})/(1+\delta_{\alpha\beta})$.

\subsection{\bf Evaluation of hadronic matrix elements at low energies} \label{bosonization}

In treating the kaon decay amplitudes, the contributions of the operators generated by NP above the EW scale are to be evaluated at the low energy of interest. This entails dealing with the QCD RG running of the Wilson coefficients by resumming the large logarithms due to the ratio of the EW scale to the chiral-symmetry breaking scale, $\Lambda_\chi$, the running between the NP and EW scales having been neglected.
We can use the one-loop QCD running results of ref.\,\cite{Liao:2019gex},
\begin{eqnarray}\nonumber
\mu{d \over d\mu}
\begin{pmatrix}
C_{1}^{uxyu^{\vphantom{|}}} \medskip \\
\tilde{C}_{1}^{uxyu}
\end{pmatrix}
&=&
-{\alpha_s \over 2\pi}
\begin{pmatrix} \displaystyle
-{3 \over N} & 0 \medskip \\
3 & \displaystyle 6 C_F
\end{pmatrix}
\begin{pmatrix}
C_{1}^{uxyu^{\vphantom{|}}} \medskip \\
\tilde{C}_{1}^{uxyu}
\end{pmatrix} ,
\\
\mu{d \over d\mu}
\begin{pmatrix}
C_{3}^{ddxy^{\vphantom{|}}} \medskip \\
\tilde{C}_{3}^{ddxy}
\end{pmatrix}
&=&
-{\alpha_s \over 2\pi}
\begin{pmatrix} \displaystyle
{2\over N}^{\vphantom{|}}+6C_F-4 & \displaystyle {2\over N}-4C_F+2 \medskip \\ \displaystyle
{4\over N}-2  & \displaystyle -{2\over N}-2C_F-2
\end{pmatrix}
\begin{pmatrix}
C_{3}^{ddxy^{\vphantom{|}}} \medskip \\
\tilde{C}_{3}^{ddxy}
\end{pmatrix} ,
\end{eqnarray}
where in the superscripts $xy=sd$ or $ds$, the color number $N=3$, and $C_F=(N^2-1)/(2N)=4/3$ is the second Casimir invariant of the color group SU(3)$_C$.
Taking the EW scale to be the $W$-boson mass $m_W^{}$, from the solutions to these RG equations~\cite{Liao:2019gex} we arrive at
\begin{align}
C_{1}^{uxyu}(\Lambda_\chi) =& 0.88\, C_{1}^{uxyu}(m_W) \,,
\nonumber \\
\tilde{C}_{1}^{uxyu}(\Lambda_\chi) =& 2.74\, \tilde{C}_{1}^{uxyu}(m_W)
+ 0.62\, C_{1}^{uxyu}(m_W) \,,
\nonumber \\
C_{3}^{ddxy}(\Lambda_\chi) =& 1.82\, C_{3}^{ddxy}(m_W)
- 0.34\, \tilde{C}_{3}^{ddxy}(m_W) \,,
\nonumber \\
\tilde{C}_{3}^{ddxy}(\Lambda_\chi)=& 0.52\, \tilde{C}_{3}^{ddxy}(m_W)
- 0.08\, C_{3}^{ddxy}(m_W) \,. \label{coeff}
\end{align}

In the matching to chiral perturbation theory ($\chi$PT), the neutrino bilinear in a dim-9 operator  behaves as a fixed external source.
Thus, we only have to work with the quark portion of the operator.
Suppose the latter has been decomposed into a sum of irreducible representations of the chiral group SU$(3)_L\times{\rm SU}(3)_R$ under which the quarks transform as
\begin{align}
q_{L,\textsf a}^{} & \,\to\, \hat L_{\textsf{ap}}^{}\, q_{L,\textsf p}^{} \,, &
\overline{q_{R,\textsf c}^{}} & \,\to\, \overline{q_{R,\textsf p}^{}}\, \hat R^\dagger_{\textsf{pc}} \,, &
q_{R,\textsf a}^{} & \,\to\, \hat R_{\textsf{ap}}^{}\, q_{R,\textsf p}^{} \,, &
\overline{q_{L,\textsf c}^{}} & \,\to\, \overline{q_{L,\textsf p}^{}}\, \hat L^\dagger_{\textsf{pc}} \,, ~~
\end{align}
where the indices $\textsf a,\textsf c,\textsf p=1,2,3$ refer to the flavor space, summation over $\textsf p$ is implicit, $\hat L\in{\rm SU}(3)_L$, and $\hat R\in{\rm SU}(3)_R$.
Given that an irreducible representation has the general form
\begin{eqnarray}
O_q=T_{\textsf{cd},\textsf{ab}}^{} \big(\overline{q_{\chi_1,\textsf c}^{}}\, \Gamma_1^{}  q_{\chi_2,\textsf a}^{}\big) \big(\overline{q_{\chi_3,\textsf d}^{}}\, \Gamma_2^{} q_{\chi_4,\textsf b}^{}\big) \,,
\end{eqnarray}
where the $T_{\textsf{cd},\textsf{ab}}$ represent pure numbers which depend on the irrep under consideration, the flavor indices $\textsf a,\textsf b,\textsf c,\textsf d=1,2,3$ are summed over, $\chi_{1,2,3,4}=L,R$, and $\Gamma_{1,2}$ designate combinations of Dirac matrices, then
promoting $T$ to be a spurion field that transforms properly together with the chiral transformations
of quarks would render $O_q$ chirally invariant.

On the $\chi$PT side, we introduce the standard matrices for the lightest octet of pseudoscalar mesons,
\begin{align}
\Sigma & = \xi^2 \,, & \xi & = \exp\left(\frac{i\Pi}{\sqrt{2}F_0}\right) , &
\Pi & = \begin{pmatrix} \displaystyle
\frac{\pi^0}{\sqrt{2}}+\frac{\eta}{\sqrt{6}} & \pi^+ & K^+
\\
\pi^- & \displaystyle \displaystyle -\frac{\pi^0}{\sqrt{2}}+\frac{\eta}{\sqrt{6}} & K^0
\\
K^- & \bar{K}^0 & \displaystyle -\sqrt{\frac{2}{3}}\, \eta
\end{pmatrix}, ~~~
\end{align}
where $F_0=F_\pi/1.0627\approx 87~\rm MeV$ is the meson decay constant in the chiral limit.
Under chiral transformations the $\Sigma$ and $\xi$ matrices transform as
\begin{align}
\Sigma & \to \hat L\Sigma\hat R^\dagger \,, & \xi & \to \hat L\xi\hat U^\dagger = \hat U\xi\hat R^\dagger \,,
\end{align}
and so
$\hat U\in{\rm SU}(3)_V$ depends on the meson fields.
From the second formula, in terms of matrix elements we have
\begin{align} \label{xixi+}
\xi_{\textsf{ab}}^{} & \to \hat L_{\textsf{ap}}^{}(\xi\hat U^\dagger)_{\textsf{pb}}^{} = (\hat U\xi)_{\textsf{ap}}^{}(\hat R^\dagger)_{\textsf{pb}}^{} \,, &
(\xi^\dagger)_{\textsf{ab}}^{} & \to \hat R_{\textsf{ap}}^{}(\xi^\dagger\hat U^\dagger)_{\textsf{pb}}^{}
= (\hat U\xi^\dagger)_{\textsf{ap}}^{}(\hat L^\dagger)_{\textsf{pb}}^{} \,. &
\end{align}
How to construct the leading-order (LO) mesonic interactions from the quark operators has been prescribed previously in the literature \cite{Graesser:2016bpz,Cirigliano:2017ymo,Liao:2019gex}.
Accordingly, eq.\,(\ref{xixi+}) implies that the matching to $\chi$PT involves
the substitutions~\cite{Liao:2019gex}
\begin{align}
\overline{q_{L,\textsf a}^{}} & \,\Rightarrow\, \xi^\dagger_{\varrho\textsf a} \,, &
q_{L,\textsf a}^{} & \,\Rightarrow\, \xi_{\textsf a\varrho}^{} \,, &
\overline{q_{R,\textsf a}^{}} & \,\Rightarrow\, \xi_{\varrho\textsf a}^{} \,, &
q_{R,\textsf a}^{} & \,\Rightarrow\, \xi^\dagger_{\textsf a\varrho} \,, & \label{rep1}
\end{align}
where the free indices $\varrho$ are to be contracted when forming an operator with $T_{\textsf{cd},\textsf{ab}}$.

For a given quark operator, the first step in the matching is to decompose it according to the irreducible representations (irreps) of the chiral group SU(3)$_L\times$SU(3)$_R$. One can easily see that ${\texttt O}_1^{usdu}$, $\tilde{\texttt O}{}_1^{usdu}$, ${\texttt O}_1^{udsu}$, and $\tilde{\texttt O}{}_1^{udsu}$ in eq.\,(\ref{opebasis2}) belong to the $8_L$$\times$$8_R$ irrep of the chiral group, while ${\texttt O}_3^{ddds}$, $\tilde{\texttt O}{}_3^{ddds}$, ${\texttt O}_3^{ddsd}$, and $\tilde{\texttt O}{}_3^{ddsd}$ belong to the $6_L$$\times$$\bar6_R$.  Then, after applying eq.\,(\ref{rep1}), one associates with the mesonic counterpart of each irrep a low energy constant $g_{\rm irrep}^{}$, which encodes nonperturbative QCD effects and is to be determined usually by fitting to data or a model calculation.

For example, the LEC associated with ${\texttt O}_1^{usdu}=(\overline{u_L}\gamma_\mu s_L)(\overline{d_R}\gamma^\mu u_R)J$, which transforms as the $8_L$$\times$$8_R$ under the chiral group, can be called $g_{8\times8}^{}$, and the leading-order chiral realization of ${\texttt O}_1^{usdu}$ is derived via the procedure
\begin{align}\nonumber
{\texttt O}_1^{usdu} & \Rightarrow\, \frac{F_0^4}{4} g_{8\times8}^{}\, \xi^\dagger_{\vartheta1} \xi_{3\varrho}^{} \xi_{\varsigma2}^{} \xi^\dagger_{1\varphi}\, \delta_{\vartheta\varphi}^{} \delta_{\varrho\varsigma}^{}\, J
\,=\, {F_0^4\over 4} g_{8\times8}^{}\, (\xi\xi)_{32}(\xi^\dagger\xi^\dagger)_{11} J
= {F_0^4\over 4} g_{8\times8}^{}\, \Sigma_{32}\Sigma^\dagger_{11} J
\\ & \Rightarrow\, \frac{g_{8\times8}^{}}{4} F_0^2 \Bigg({3 \over\sqrt{2}}\pi^0\bar{K}^0-\pi^+K^-\Bigg)J
+ \frac{i g_{8\times8}^{}}{4\sqrt2} F_0^3\bar{K}^0 J \,+\,\cdots \,, \label{o1usdu}
\end{align}
where $F_0^4/4$ is a normalization factor~\cite{Cirigliano:2017ymo}, other contractions among $\varepsilon,\varrho,\varsigma,\varphi$ in the first line vanish due to the unitarity of $\xi$, and the ellipsis stands for terms with the $\eta$ field and more than two meson fields.
This result is independent of the Lorentz and color structures of ${\texttt O}_1^{usdu}$ and follows from its transformation properties as an irrep of the chiral group.
Evidently, $g_{8\times8}^{}$ has mass dimension two.
This example can be understood from the perspective of bosonization of each quark bilinear: Fierz-transforming ${\cal Q}_1^{usdu}$ yields \,$-2\bar u_L^{\textsf m}u_R^{\textsf n}\bar s_L^{\textsf n}d_R^{\textsf m}$\, which is summed over the color labels $\textsf m,\textsf n$ and has a lower-order chiral realization than $(\overline{u_L}\gamma_\mu s_L)(\overline{d_R}\gamma^\mu u_R)$ being naively taken to correspond to the higher order $(\partial^\mu\Sigma\Sigma^\dagger)_{31} (\Sigma^\dagger\partial_\mu\Sigma)_{12}$ due to the derivatives. If the operator comprised instead solely left-handed quarks, $(\overline{u_L}\gamma_\mu s_L)(\overline{d_L}\gamma^\mu u_L)$, its chiral realization would have to be of the form $(\partial^\mu\Sigma\Sigma^\dagger)_{op} (\partial_\mu\Sigma\Sigma^\dagger)_{xy}$ because no scalar density can be constructed with such quarks.

In table\,\,\ref{tab1} we provide the chiral realization of the quark component of each of the operators in eq.\,(\ref{opebasis2}) according to their chiral irreps. In the last column, we display the leading-order contributions to the $K\to\pi$ transitions in terms of the mesonic operators
\begin{align} \label{OS}
{\cal Q}_{1/2}^S & =\, F_0^2 \left( K^+\pi^--{1\over \sqrt{2}} K^0\pi^0 \right) , &
{\cal Q}_{3/2}^S & =\, F_0^2 \left( K^+\pi^-+\sqrt{2} K^0\pi^0 \right) ,
\end{align}
or their Hermitian conjugates, which correspond to definite isospin changes $\Delta I=1/2$ and 3/2, respectively.\footnote{For completeness, in appendix\,\,\ref{isodec} we give the decomposition of the quark part of each of the operators in eq.\,(\ref{opebasis2}) in terms of their $\Delta I=1/2,3/2$ components.}
It is worth noting that in the combination $5{\cal Q}_{1/2}^S-2{\cal Q}_{3/2}^S$, which occurs in all of the lines in table\,\,\ref{tab1} and hence also implicitly in eq.\,(\ref{o1usdu}), the size of the $K^0\pi^0$ term relative to the $K^+\pi^-$ one is three times that in ${\cal Q}_{1/2}^S$ alone. It follows that every one of these operators can potentially break the GN relation.

The table also shows that each operator with a tilde and its counterpart without it have the same chiral realization but their LECs are different. This is attributable to the fact that the chiral realization of a quark-level operator relies only on its representation under the chiral group, whereas the LEC encodes QCD effects. Numerically, we adopt the values of the LECs extracted from ref.\,\cite{Cirigliano:2017ymo} which employed $\chi$PT to connect the matrix elements of $\pi^+\to\pi^-$ transitions to kaon-mixing matrix elements for which lattice QCD results were available.
Thus, we have
\begin{align}
g_{8\times8}^{} & \,=\, -2.9{\rm~GeV}^2 \,, & \tilde g_{8\times8}^{} & \,=\, -12.4{\rm~GeV}^2 \,,
\nonumber \\
g_{6\times\bar6}^{} & \,=\, 2.7{\rm~GeV}^2 \,, & \tilde g_{6\times\bar6}^{} & \,=\, -0.91{\rm~GeV}^2 \,. ~~~ \label{LEC1}
\end{align}

\begin{table}
\begin{tabular}{| c | c | c | c |}
\hline
~Operator~ & ~Chiral irrep~ & ~Chiral realization~ & ~Contributions to $K\to\pi$~
\\ \hline\hline
${\texttt O}_1^{usdu} \vphantom{\Big|_|^|}$ & $8_L\times8_R$  &${1\over 4}F_0^2 g_{8\times8}^{}\Sigma_{32}\Sigma^\dagger_{11}$ &
~$-{1\over 12}g_{8\times8}^{} \left( 5{\cal Q}_{1/2}^{S\dagger}-2{\cal Q}_{3/2}^{S\dagger} \right)$~
\\
$\tilde{\texttt O}{}_1^{usdu} \vphantom{\Big|_|^|}$ & $8_L\times8_R$ &${1\over 4}F_0^2\tilde g_{8\times8}^{}\Sigma_{32}\Sigma^\dagger_{11}$ &
$-{1\over 12}\tilde g_{8\times8}^{} \left( 5{\cal Q}_{1/2}^{S\dagger}-2{\cal Q}_{3/2}^{S\dagger} \right)$
\\
${\texttt O}_1^{udsu} \vphantom{\Big|_|^|}$ & $8_L\times8_R$ &${1\over 4}F_0^2 g_{8\times8}^{}\Sigma_{23}\Sigma^\dagger_{11}$ &
$-{1\over 12}g_{8\times8}^{} \left( 5{\cal Q}_{1/2}^{S}-2{\cal Q}_{3/2}^{S}\right)$
\\
$\tilde{\texttt O}{}_1^{udsu} \vphantom{\Big|_|^|}$ & $8_L\times8_R$ &${1\over 4}F_0^2\tilde g_{8\times8}^{}\Sigma_{23}\Sigma^\dagger_{11}$ &
$-{1\over 12}\tilde g_{8\times8}^{} \left( 5{\cal Q}_{1/2}^{S}-2{\cal Q}_{3/2}^{S} \right)$
\\\hline
${\texttt O}_3^{ddds} \vphantom{\Big|_|^|}$ & $6_L\times \bar 6_R$ &${1\over 4}F_0^2 g_{6\times\bar6}^{}\Sigma_{22}\Sigma_{32}$ &
$-{1\over12}g_{6\times\bar6}^{} \left( 5{\cal Q}_{1/2}^{S\dagger}-2{\cal Q}_{3/2}^{S\dagger} \right)$
\\
$\tilde{\texttt O}{}_3^{ddds} \vphantom{\Big|_|^|}$ & $6_L\times  \bar  6_R$ &${1\over 4}F_0^2\tilde g_{6\times\bar6}^{}\Sigma_{22}\Sigma_{32}$ &
$-{1\over 12}\tilde g_{6\times\bar6}^{} \left( 5 {\cal Q}_{1/2}^{S\dagger}-2{\cal Q}_{3/2}^{S\dagger} \right)$
\\
${\texttt O}_3^{ddsd} \vphantom{\Big|_|^|}$ & $6_L\times  \bar  6_R$ &${1\over 4}F_0^2 g_{6\times \bar6}^{}\Sigma_{23}\Sigma_{22}$ &
$-{1\over 12}g_{6\times\bar6}^{} \left( 5 {\cal Q}_{1/2}^{S}-2{\cal Q}_{3/2}^{S} \right)$
\\
$\tilde{\texttt O}{}_3^{ddsd} \vphantom{\Big|_|^|}$ & $6_L\times\bar6_R$ &${1\over 4}F_0^2\tilde g_{6\times\bar6}^{}\Sigma_{23}\Sigma_{22}$ &
$-{1\over 12}\tilde g_{6\times\bar6}^{} \left( 5 {\cal Q}_{1/2}^{S}-2{\cal Q}_{3/2}^{S} \right)$
\\\hline
\end{tabular}
\caption{The chiral representation and realization of the four-quark portion of each of the effective dim-9 operators in eq.\,(\ref{opebasis2}) relevant to the $K\to\pi$ processes. In the last column, ${\cal Q}_{1/2}^S$ and ${\cal Q}_{3/2}^S$ are the mesonic operators defined in eq.\,(\ref{OS}) and correspond to $\Delta I=1/2$ and 3/2 transitions, respectively.}
\label{tab1}
\end{table}

\subsection{\bf Numerical analysis}

From the results of the preceding subsection, we can write down the effective interactions pertinent to $K^+\to\pi^+\nu\nu,\pi^+\bar\nu\bar\nu$  and $K_L\to\pi^0\nu\nu,\pi^0\bar\nu\bar\nu$, namely
\begin{eqnarray}\nonumber
C_{1}^{usdu}{\texttt O}^{usdu}_1+{\text{H.c.}} &\Rightarrow&
 {g_{8\times8}^{}\over 4}F_0^2 \bigg[ \frac{3}{2} \Big(C_{1}^{usdu} J+ C_{1}^{usdu*} J^\dagger\Big) \pi^0K_L-C_{1}^{usdu*} J^\dagger\pi^-K^+ \bigg] \,,
\\\nonumber
C_{1}^{udsu}{\texttt O}^{udsu}_1+{\text{H.c.}} &\Rightarrow&
{g_{8\times8}^{}\over 4}F_0^2 \bigg[  \frac{3}{2} \Big(C_{1}^{udsu} J+ C_{1}^{udsu*} J^\dagger\Big) \pi^0K_L-C_{1}^{udsu}J\pi^-K^+ \bigg] \,,
\\\nonumber
C_{3}^{ddds} {\texttt O}^{ddds}_3+{\text{H.c.}} &\Rightarrow&
 {g_{6\times6}^{}\over 4}F_0^2 \bigg[ \frac{3}{2} \Big(C_{3}^{ddds} J+C_{3}^{ddds*} J^\dagger\Big) \pi^0K_L-C_{3}^{ddds*}J^\dagger\pi^-K^+ \bigg] \,,
\\\label{effint}
C_{3}^{ddsd}{\texttt O}^{ddsd}_3+{\text{H.c.}} &\Rightarrow&
{g_{6\times6}^{}\over 4}F_0^2 \bigg[  \frac{3}{2} \Big(C_{3}^{ddsd}J+ C_{3}^{ddsd*}J^\dagger\Big) \pi^0K_L-C_{3}^{ddsd}J\pi^-K^+ \bigg] \,,
\end{eqnarray}
and analogous expressions for the tilded operators, where $J^\dagger=(\overline{\nu_\alpha}\nu_\beta^{\textsc c})/(1+\delta_{\alpha\beta})$.
With these, we arrive at the decay amplitudes
\begin{align}\nonumber
\mathcal{A}_{K_L\to \pi^0\nu_\alpha\nu_\beta} & \,=\, \frac{3}{8}\, F_0^2\, \big(C_A^*+C_B^*\big)\, \overline{\nu_\alpha^{}}\nu_\beta^{\textsc c} \,, &
\mathcal{A}_{K^+\to \pi^+\nu_\alpha\nu_\beta} & \,=\, -\frac{1}{4}\, F_0^2\, C_B^*\, \overline{\nu_\alpha^{}}\nu_\beta^{\textsc c} \,,
\\\label{amp}
\mathcal{A}_{K_L\to \pi^0\bar\nu_\alpha\bar\nu_\beta} & \,=\, \frac{3}{8}\, F_0^2\, \big( C_A^{}+C_B^{}\big)\, \overline{\nu_\alpha^{\textsc c}}\nu_\beta^{} \,, &
\mathcal{A}_{K^+\to \pi^+\bar\nu_\alpha\bar\nu_\beta} & \,=\, -\frac{1}{4}\, F_0^2\, C_A^{}\, \overline{\nu_\alpha^{\textsc c}}\nu_\beta^{} \,,
\end{align}
involving the effective coupling constants
\begin{align} \label{CA}
C_A & \,=\, g_{8\times8}^{} C_{1}^{udsu}(\Lambda_\chi) + \tilde g_{8\times8}^{}\tilde{C}_{1}^{udsu}(\Lambda_\chi) + g_{6\times6}^{} C_{3}^{ddsd}(\Lambda_\chi) + \tilde g_{6\times6}\tilde{C}_{3}^{ddsd} (\Lambda_\chi) \,,
\nonumber \\
C_B & \,=\, g_{8\times8}^{}C_{1}^{usdu}(\Lambda_\chi) + \tilde g_{8\times8}^{} \tilde{C}_{1}^{usdu}(\Lambda_\chi) + g_{6\times6^{} }C_{3}^{ddds}(\Lambda_\chi) + \tilde g_{6\times6}^{} \tilde{C}_{3}^{ddds} (\Lambda_\chi) \,, ~~
\end{align}
which implicitly carry the neutrino family labels $\alpha$ and $\beta$.
From eq.\,(\ref{amp}), we obtain the spin-summed absolute squares
\begin{align}\nonumber
\mbox{\footnotesize$\displaystyle\sum_{\rm spins}$}\, |\mathcal{A}_{K_L\to \pi^0\nu_\alpha\nu_\beta}|^2 & \,=\, \frac{9}{32}\, F_0^4 \big|C_A+C_B\big|^2\hat s\,, &
\mbox{\footnotesize$\displaystyle\sum_{\rm spins}$}\, |\mathcal{A}_{K^+\to \pi^+\nu_\alpha\nu_\beta}|^2 & \,=\, \frac{1}{8}\, F_0^4 |C_B|^2\hat s\,,
\\
\mbox{\footnotesize$\displaystyle\sum_{\rm spins}$}\, |\mathcal{A}_{K_L\to \pi^0\bar\nu_\alpha\bar\nu_\beta}|^2 & \,=\, \frac{9}{32}\, F_0^4\, \big|C_A+C_B\big|^2\hat s\,,  &
\mbox{\footnotesize$\displaystyle\sum_{\rm spins}$}\, |\mathcal{A}_{K^+\to \pi^+\bar\nu_\alpha\bar\nu_\beta}|^2 & \,=\, \frac{1}{8}\, F_0^4\, |C_A|^2\hat s\,,
\end{align}
where $\hat s=(p_1+p_2)^2$. The $\nu_\alpha\nu_\beta$ and $\bar\nu_\alpha\bar\nu_\beta$ channels having no interference with each other, their branching fractions add up to
\begin{align}
\mathcal{B}(K_L\to\pi^0\nu_\alpha\nu_\beta)+\mathcal{B}(K_L\to\pi^0\bar\nu_\alpha\bar\nu_\beta) & \,=\, {9F_0^4\over16} {|C_A+C_B|^2\over 1+\delta_{\alpha\beta} }{ \tau_{K_L} \over 2m_{K^0}}\int d\Pi_3\hat s
\nonumber \\ & \,=\, 1.42\times10^{-9}{|\hat{C}_A+\hat{C}_B|^2 \over 1+\delta_{\alpha\beta}} \,,
\nonumber \\
\mathcal{B}(K^+\to\pi^+\nu_\alpha\nu_\beta)+\mathcal{B}(K^+\to\pi^+\bar\nu_\alpha\bar\nu_\beta) & \,=\, {F_0^4\over8}{\big(|C_A|^2+|C_B|^2\big)\over1+\delta_{\alpha\beta}}{\tau_{K^+} \over 2m_{K^+}}\int d\Pi_3\hat s
\nonumber \\ & \,=\, 6.98\times10^{-11}{ |\hat{C}_A|^2+|\hat{C}_B|^2\over 1+\delta_{\alpha\beta}} \,,
\label{BK}
\end{align}
where $\tau_{K_L}$ and $\tau_{K^+}$ stand for the measured $K_L$ and $K^+$ lifetimes~\cite{Tanabashi:2018oca}, the factor $1/(1+\delta_{\alpha\beta})$ in each equation accounts for the identical particles in the final state if $\alpha=\beta$, and $d\Pi_3$ denotes the three-body phase-space factor. Also, in eq.\,(\ref{BK}) we have defined the dimensionless parameters
\begin{align}\nonumber
\hat{C}_A & \,=\, (50{\rm\;GeV})^5 \Big[ C_{1}^{udsu}(m_W) + 3.3 \tilde{C}_{1}^{udsu}(m_W)
-0.55C_{3}^{ddsd}(m_W) + 0.15\tilde{C}_{3}^{ddsd}(m_W) \Big] \,,
\\
\hat{C}_B & \,=\, (50{\rm\;GeV})^5 \Big[ C_{1}^{usdu}(m_W) + 3.3\tilde{C}_{1}^{usdu}(m_W)
- 0.55C_{3}^{ddds}(m_W) + 0.15\tilde{C}_{3}^{ddds}(m_W) \Big] \,,
\end{align}
which incorporate eq.\,(\ref{coeff}) for the RG-running effects on the parameters between $\mu=\Lambda_\chi$ and $\mu=m_W$ and eq.\,\,(\ref{LEC1}) for the LEC values.

Assuming that the operators involve only one pair of $\alpha$ and $\beta\neq\alpha$,
since these modes do not interfere with the SM ones, we can combine eq.\,(\ref{BK}) with their SM counterparts to find
\begin{align}\nonumber
\mathcal{B}(K_L\to\pi^0\mbox{+}E_{\rm miss}) & \,=\, \mathcal{B}(K_L\to\pi^0\nu\bar\nu)_{\rm SM}^{} + \mathcal{B}(K_L\to\pi^0\nu_\alpha\nu_\beta) + \mathcal{B}(K_L\to\pi^0\bar\nu_\alpha\bar\nu_\beta) \,,
\\ \label{K->pi+inv}
\mathcal{B}(K^+\to\pi^+\mbox{+}E_{\rm miss}) & \,=\, \mathcal{B}(K^+\to\pi^+\nu\bar\nu)_{\rm SM}^{} + \mathcal{B}(K^+\to\pi^+\nu_\alpha\nu_\beta) + \mathcal{B}(K^+\to\pi^+\bar\nu_\alpha\bar\nu_\beta) \,.
\end{align}
Accordingly, the definition of $r_{\cal B}$ should be modified, and we now have
\begin{align}
r_{\cal B} & \,=\, {\mathcal{B}(K_L\to\pi^0\mbox{+}E_{\rm miss}) \over
\mathcal{B}(K^+\to\pi^+\mbox{+}E_{\rm miss})}
=r_{\cal B}^{\rm SM}+{(r_{\cal B}^{\rm NP}-r_{\cal B}^{\rm SM})\epsilon \over 1+\epsilon}
\leq0.36+{40.3\epsilon \over 1+\epsilon} \,,
\end{align}
where
\begin{align}
r_{\cal B}^{\rm NP} & \,=\, \frac{ {\cal B}(K_L\to\pi^0\nu_\alpha\nu_\beta) + {\cal B}(K_L\to\pi^0\bar\nu_\alpha\bar\nu_\beta) }{ {\cal B}(K^+\to\pi^+\nu_\alpha\nu_\beta) + {\cal B}(K^+\to\pi^+\bar\nu_\alpha\bar\nu_\beta) } \,=\, \frac{20.3\, \big|\hat C_A+\hat C_B\big|^2}{\big|\hat C_A\big|^2 + \big|\hat C_B\big|^2} \,\leq\,  40.6 \,,
\nonumber \\
\epsilon & \,=\, {\mathcal{B}(K^+\to\pi^+\nu_\alpha\nu_\beta) + \mathcal{B}(K^+\to\pi^+\bar\nu_\alpha\bar\nu_\beta) \over \mathcal{B}(K^+\to\pi^+\nu\bar\nu)_{\rm SM}^{}} \,.
\end{align}
The most recent measurement ${\mathcal B}(K^+\to\pi^+\nu\bar\nu)_{\rm NA62} < 1.85\times 10^{-10}$ at 90\% CL~\cite{na62} entails that $\epsilon\,\raisebox{1pt}{\footnotesize$\lesssim$}\,1$ and consequently $r_{\cal B}$ is capped to be about $20.5$.
This can accommodate KOTO's anomalous events~\cite{koto}.

\begin{figure}[t]
\includegraphics[width=11cm]{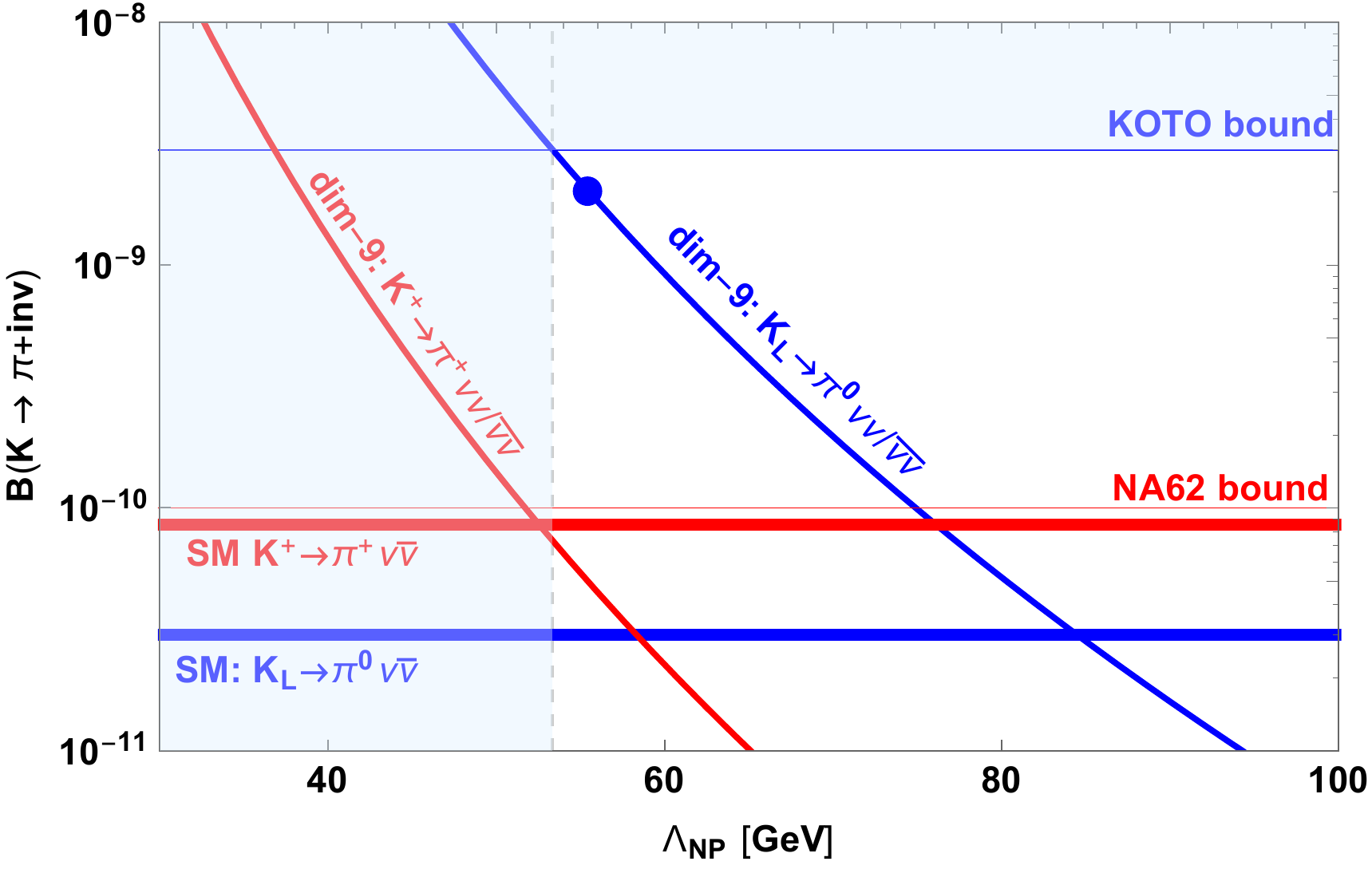} \vspace{-1ex}
\caption{The branching fractions of $K\to\pi\nu\nu,\pi\bar\nu\bar\nu$ arising from the dim-9 operators versus the NP scale $\Lambda_{\textsc{np}}$ for the example described in the text. Also displayed are the corresponding SM predictions for  $K\to\pi\nu\bar\nu$ (red and blue horizontal bands) and upper limits from KOTO \cite{Ahn:2018mvc} and NA62 \cite{na62} (blue and red horizontal thin lines). The light-blue region is excluded by the KOTO bound. The blue dot corresponds to KOTO's three events.}
\label{Ktopinunu}
\end{figure}

Imposing on eq.\,(\ref{K->pi+inv}) the KOTO 15~\cite{Ahn:2018mvc} and NA62~\cite{na62} limits and further assuming that the only nonvanishing coefficients are $C_{1}^{udsu}(m_W)=C_{1}^{usdu}(m_W)=\Lambda_{\textsc{np}}^{-5}$, we have
\begin{align}\nonumber
\mathcal{B}(K_L\to\pi^0\mbox{+}E_{\rm miss}) & \,=\,
3.0 \times 10^{-11} + 5.7\times 10^{-9}\left({50~\rm GeV \over \Lambda_{\textsc{np}}}\right)^{10}
\,\leq\, 3.0\times 10^{-9} \,,
\\ \label{dim9k2pnn}
\mathcal{B}(K^+\to\pi^+\mbox{+}E_{\rm miss}) & \,=\,
8.5\times10^{-11} + 1.4\times10^{-10}\left({50~\rm GeV \over \Lambda_{\textsc{np}}}\right)^{10}
\,\leq\, 1.85\times 10^{-10} \,,
\end{align}
where the first number in each line is the corresponding SM central value~\cite{Charles:2004jd,ckmfitter,Tanabashi:2018oca}. The stronger of the empirical limits, from KOTO in the first line, translates into $\Lambda_{\textsc{np}}\,\raisebox{1pt}{\footnotesize$\gtrsim$}\,53$\,GeV. In figure\,\,\ref{Ktopinunu} we illustrate the dependence of $\mathcal{B}(K_L\to\pi^0+E_{\rm miss})$ and $\mathcal{B}(K^+\to\pi^++E_{\rm miss})$ in eq.\,(\ref{dim9k2pnn}) on $\Lambda_{\textsc{np}}$. Also plotted are the corresponding SM predictions and limits from KOTO~\cite{Ahn:2018mvc} and NA62~\cite{na62}.
We see that the NP needs to have an effective scale $\Lambda_{\textsc{np}}={\cal O}$(60\,\,GeV) if it is to be responsible for the KOTO anomaly. The preferred $\Lambda_{\textsc{np}}$ is significantly below the EW scale $\sim v\simeq246\;$GeV,\, which implies that for \,$\Lambda_{\textsc{np}}\sim v$\, the dim-9 operators would have negligible impact on \,$K\to\pi$+$E_{\rm miss}$\, and thus respect the GN bound.

If we repeat the above steps with the dim-10 operators discussed earlier, a scaling factor of $v/\Lambda_{\textsc{np}}$ will accompany them.
As mentioned in subsection\,\,\ref{k2p2n}, they conserve lepton number and hence, if not flavor-violating, interfere with the SM contribution.
For $\Lambda_{\textsc{np}}<v$, this scaling factor helps raise $\Lambda_{\textsc{np}}$ slightly, but the latter still will not be very close to the EW scale, {\footnotesize$\sim$}\,$v$, if the NP presence in \,$K\to\pi$+$E_{\rm miss}$\, is to be within the current experimental sensitivity reaches.
Hence for $\Lambda_{\textsc{np}}\,\raisebox{1pt}{\footnotesize$\gtrsim$}\,v$ the dim-10 operators would also have very little influence on these modes.

\section{GN-bound violation via EFT operators for \boldmath$K\to\pi S{\rm~or~}\pi SS$\label{k2ps}}

In the preceding section, the problem of the NP scale $\Lambda_{\textsc{np}}$ being too low can be ascribed to the high dimension of the SMEFT operators.
As sketched in section\,\,\ref{intro}, any NP that can induce $K_L\to\pi^0\cal X$ with a rate exceeding the SM expectation without conflicting with the $K^+$ data must have an effective scale $\Lambda_{\textsc{np}}$ which goes roughly as $\big(23000{\rm\,TeV}^2\,m_K^{\textsc n}\big){}^{1/(2+\textsc n)}$, where $m_K$ is the kaon mass and $\textsc n$ the mass dimension of the field content of $\cal X$.
Therefore, one way to increase $\Lambda_{\textsc{np}}$ is by reducing the dimension of the operators, which is 6+{\textsc n}.
If in the dim-9 operators examined above the neutrino pair, which has $\textsc n=3$, is replaced with a scalar field $S$, which has $\textsc n=1$, the dimension of the operators can be decreased by 2 and in turn $\Lambda_{\textsc{np}}$ can be raised to the TeV level.
If ${\cal X}=SS$ instead, the scale will become $\Lambda_{\textsc{np}}={\cal O}(v)$.
In this section, we explore how SMEFT four-quark operators supplemented with $S$, which we take to be real and a singlet under the SM gauge group, can give rise to $K\to\pi S$ transitions which break the GN bound.
Moreover, we apply a similar treatment to the $K\to\pi SS$ case.

\subsection{\bf Operators and matching}

As is clear from the last paragraph, SMEFT operators that directly give rise to $K\to\pi S$ at leading order have to be of dimension seven (dim-7).
Since $S$ is a SM-gauge singlet, the quark portions of these operators are none other than the SMEFT dimension-six (dim-6) four-quark operators \cite{Buchmuller:1985jz,Grzadkowski:2010es} which contribute to nonleptonic kaon decays.
In the first column of table\,\,\ref{smeftleft}, we list these dim-6 operators in the Warsaw basis \cite{Grzadkowski:2010es}. In the middle column, we exhibit the relevant operators with one $s$-quark field in the LEFT approach~\cite{Jenkins:2017jig}.
The third column contains the results for the Wilson coefficients at the electroweak scale from the matching of the SMEFT operators onto the LEFT operators~\cite{Jenkins:2017jig}.
For the $K\to\pi S$ and $K\to\pi SS$ transitions, we just multiply all those operators by an $S$ field and a pair of them, respectively, and the matched Wilson coefficients should be understood as new parameters associated with the operators.

\begin{table}[t]
\begin{tabular}{|l | l | l | }
\hline SMEFT operators & LEFT operators & Matching at the EW scale
\\ \hline\hline
$Q_{qq}^{(1)}=(\bar Q\gamma_\mu Q)(\bar Q\gamma_\mu Q)$ &
$\textsl{\texttt O}^{V,LL}_{ddsd}=(\overline{d_L}\gamma_\mu d_L)(\overline{s_L}\gamma^\mu d_L)$ &
${\tt C}^{V,LL}_{ddsd}=C_{qq}^{(1),1121}+C_{qq}^{(3),1121} \vphantom{\Big|^|}$
\\
$Q_{qq}^{(3)}=(\bar Q \tau^I\gamma_\mu  Q)(\bar Q\tau^I\gamma^\mu Q)$ &
$\textsl{\texttt O}^{V1,LL}_{uusd}=(\overline{u_L}\gamma_\mu u_L)(\overline{s_L}\gamma^\mu d_L)$ &
${\tt C}^{V1,LL}_{uusd}=C_{qq}^{(1),1121}-\frac{1}{3}C_{qq}^{(3),1121} \vphantom{\Big|_1}$~
\\
& $\textsl{\texttt O}^{V8,LL}_{uusd}=(\overline{u_L}T^A\gamma_\mu u_L)(\overline{s_L}T^A\gamma^\mu d_L)$  &
${\tt C}^{V8,LL}_{uusd}=4C_{qq}^{(3),1121} \vphantom{\Big|_1}$
\\\hline
$Q_{dd}=(\bar d\gamma_\mu d)(\bar d\gamma^\mu d)$ &
$\textsl{\texttt O}^{V,RR}_{ddsd}=(\overline{d_R}\gamma_\mu d_R)(\overline{s_R}\gamma^\mu d_R)$ &
${\tt C}^{V,RR}_{ddsd}=C_{dd}^{1121} \vphantom{\Big|^|}$
\\
$Q_{ud}^{(1)}=(\bar u \gamma_\mu u)(\bar d \gamma^\mu  d)$ &
$\textsl{\texttt O}^{V1,RR}_{uusd}=(\overline{u_R}\gamma_\mu u_R)(\overline{s_R}\gamma^\mu d_R)$ &
${\tt C}^{V1,RR}_{uusd}=C_{ud}^{(1),1121} \vphantom{\Big|_1}$
\\
$Q_{ud}^{(8)}=(\bar uT^A \gamma_\mu u)(\bar d T^A\gamma^\mu  d)$ &
$\textsl{\texttt O}^{V8,RR}_{uusd}=(\overline{u_R}T^A\gamma_\mu u_R)(\overline{s_R}T^A\gamma^\mu d_R)$~ &
${\tt C}^{V8,RR}_{uusd}=C_{ud}^{(8),1121} \vphantom{\Big|_1}$
\\\hline
$Q_{qu}^{(1)}=(\bar Q\gamma_\mu Q)(\bar u\gamma^\mu u)$ &
$\textsl{\texttt O}^{V1,LR}_{sduu}=(\overline{s_L}\gamma_\mu d_L)(\overline{u_R}\gamma^\mu u_R)$ &
${\tt C}^{V1,LR}_{sduu}=C_{qu}^{(1),2111} \vphantom{\Big|^|}$
\\
$Q_{qu}^{(8)}=(\bar QT^A\gamma_\mu Q)(\bar u T^A \gamma^\mu u)$ &
$\textsl{\texttt O}^{V8,LR}_{sduu}=(\overline{s_L}T^A\gamma_\mu d_L)(\overline{u_R}T^A\gamma^\mu u_R)$ &
${\tt C}^{V8,LR}_{sduu}=C_{qu}^{(8),2111} \vphantom{\Big|_1}$
\\
$Q_{qd}^{(1)}=(\bar Q\gamma_\mu Q)(\bar d\gamma^\mu d)$ &
$\textsl{\texttt O}^{V1,LR}_{sddd}=(\overline{s_L}\gamma_\mu d_L)(\overline{d_R}\gamma^\mu d_R)$ &
${\tt C}^{V1,LR}_{sddd}=C_{qd}^{(1),2111} \vphantom{\Big|_1}$
\\
$Q_{qd}^{(8)}=(\bar Q T^A\gamma_\mu Q)(\bar d T^A\gamma^\mu d)$ &
$\textsl{\texttt O}^{V8,LR}_{sddd}=(\overline{s_L}T^A\gamma_\mu d_L)(\overline{d_R}T^A\gamma^\mu d_R)$ &
${\tt C}^{V8,LR}_{sddd}=C_{qd}^{(8),2111} \vphantom{\Big|_1}$
\\\hline
$Q_{qd}^{(1)}=(\bar Q\gamma_\mu Q)(\bar d\gamma^\mu d)$ &
$\textsl{\texttt O}^{V1,LR}_{uusd}=(\overline{u_L}\gamma_\mu u_L)(\overline{s_R}\gamma^\mu d_R)$ &
${\tt C}^{V1,LR}_{uusd}=C_{qd}^{(1),1121} \vphantom{\Big|^|}$
\\
$Q_{qd}^{(8)}=(\bar Q T^A\gamma_\mu Q)(\bar d T^A\gamma^\mu d)$ &
$\textsl{\texttt O}^{V8,LR}_{uusd}=(\overline{u_L}T^A\gamma_\mu u_L)(\overline{s_R}T^A\gamma^\mu d_R)$ &
${\tt C}^{V8,LR}_{uusd}=C_{qd}^{(8),1121} \vphantom{\Big|_1}$
\\
& $\textsl{\texttt O}^{V1,LR}_{ddsd}=(\overline{d_L}\gamma_\mu d_L)(\overline{s_R}\gamma^\mu d_R)$ &
${\tt C}^{V1,LR}_{ddsd}=C_{qd}^{(1),1121} \vphantom{\Big|_1}$
\\
& $\textsl{\texttt O}^{V8,LR}_{ddsd}=(\overline{d_L}T^A\gamma_\mu d_L)(\overline{s_R}T^A\gamma^\mu d_R)$ &
${\tt C}^{V8,LR}_{ddsd}=C_{qd}^{(8),1121} \vphantom{\Big|_1}$
\\\hline
$Q_{quqd}^{(1)}=\epsilon_{ij}(\bar Q^iu)(\bar Q^jd)$ &
$\textsl{\texttt O}^{S1,RR}_{uusd}=(\overline{u_L}u_R)(\overline{s_L}d_R)$ &
${\tt C}^{S1,RR}_{uusd}=C_{quqd}^{(1),1121} \vphantom{\Big|^|}$
\\
$Q_{quqd}^{(8)}=\epsilon_{ij}(\bar Q^i T^Au)(\bar Q^jT^A d)$~ &
$\textsl{\texttt O}^{S8,RR}_{uusd}=(\overline{u_L}T^Au_R)(\overline{s_L}T^Ad_R)$ &
${\tt C}^{S8,RR}_{uusd}=C_{quqd}^{(8),1121} \vphantom{\Big|_1}$
\\
& $\textsl{\texttt O}^{S1,RR}_{udsu}=(\overline{u_L}d_R)(\overline{s_L}u_R)$ &
${\tt C}^{S1,RR}_{uusd}=-C_{quqd}^{(1),2111} \vphantom{\Big|_1}$
\\
& $\textsl{\texttt O}^{S8,RR}_{udsu}=(\overline{u_L}T^Ad_R)(\overline{s_L}T^Au_R)$ &
${\tt C}^{S8,RR}_{uusd}=-C_{quqd}^{(8),2111} \vphantom{\Big|_1}$
\\\hline
$Q_{quqd}^{(1)\dagger}=\epsilon_{ij}(\bar u  Q^i)(\bar dQ^j)$ &
$\textsl{\texttt O}^{S1,LL}_{uusd}=(\overline{u_R}u_L)(\overline{s_R}d_L)$  &
${\tt C}^{S1,LL}_{uusd}=C_{quqd}^{(1),1112*} \vphantom{\Big|^|}$
\\
$Q_{quqd}^{(8)\dagger}=\epsilon_{ij}(\bar u T^AQ^i)(\bar dT^A Q^j)$ &
$\textsl{\texttt O}^{S8,LL}_{uusd}=(\overline{u_R}T^Au_L)(\overline{s_R}T^Ad_L)$ &
${\tt C}^{S8,LL}_{uusd}=C_{quqd}^{(8),1112*} \vphantom{\Big|_1}$
\\
& $\textsl{\texttt O}^{S1,LL}_{udsu}=(\overline{u_R}d_L)(\overline{s_R}u_L)$&
${\tt C}^{S1,LL}_{udsu}=-C_{quqd}^{(1),1112*} \vphantom{\Big|_1}$
\\
& $\textsl{\texttt O}^{S8,LL}_{udsu}=(\overline{u_R}T^Ad_L)(\overline{s_R}T^Au_L)$ &
${\tt C}^{S8,LL}_{udsu}=-C_{quqd}^{(8),1112*} \vphantom{\Big|_1}$
\\\hline
\end{tabular}
\caption{Columns 1 and 2: the SMEFT and LEFT four-quark operators contributing to $K\to\pi S(S)$. Column 3: the results for the Wilson coefficients at the electroweak scale from the matching of the former operators onto the latter.}
\label{smeftleft}
\end{table}

\begin{table}[!b] \medskip
\begin{tabular}{| l | l | }
\hline Chiral irrep & Contributions to $K\to\pi$
\\ \hline\hline
$\textsl{\texttt O}^{V,LL}_{ddsd}=\textsl{\texttt O}^{V,LL}_{ddsd}\big|_{27\times1}^{}+\textsl{\texttt O}^{V,LL}_{ddsd}\big|_{8\times1}^{} \vphantom{\Big|_|^1}$ &
${1\over 18}g_{27\times1}^{} \left( 2 {\cal Q}_{1/2}^V-5 {\cal Q}_{3/2}^V \right) + {1\over6}g_{8\times1}^{}{\cal Q}_{1/2}^V$
\\
$\textsl{\texttt O}^{V1,LL}_{uusd}=\textsl{\texttt O}^{V1,LL}_{uusd}\big|_{27\times1}^{}+\textsl{\texttt O}^{V1,LL}_{uusd}\big|_{8\times1}^{} \vphantom{\Big|_|}$ &
${1\over 18}g_{27\times1}^{} \left( {\cal Q}_{1/2}^V+5{\cal Q}_{3/2}^V\right)-\frac{1}{3}g_{8\times1}^{}{\cal Q}_{1/2}^V$
\\
$\textsl{\texttt O}^{V8,LL}_{uusd}=\textsl{\texttt O}^{V8,LL}_{uusd}\big|_{27\times1}+\textsl{\texttt O}^{V8,LL}_{uusd}\big|_{8\times1}^{} \vphantom{\Big|_|}$ &
${1\over 54}g_{27\times1}^{} \left( {\cal Q}_{1/2}^V+5{\cal Q}_{3/2}^V\right)+{11\over 36}g_{8\times1}^{}{\cal Q}_{1/2}^V$
\\ \hline
$\textsl{\texttt O}^{V,RR}_{ddsd}=\textsl{\texttt O}^{V,RR}_{ddsd}\big|_{1\times27}^{}+\textsl{\texttt O}^{V,RR}_{ddsd}\big|_{1\times8}^{} \vphantom{\Big|_|^1}$ &
${1\over 18}g_{1\times27}^{} \left( 2{\cal Q}_{1/2}^V-5{\cal Q}_{3/2}^V\right) + {1\over6}^{}g_{1\times8}^{}{\cal Q}_{1/2}^V$
\\
$\textsl{\texttt O}^{V1,RR}_{uusd}=\textsl{\texttt O}^{V1,RR}_{uusd}\big|_{1\times27}^{}+\textsl{\texttt O}^{V1,RR}_{uusd}\big|_{1\times8}^{} \vphantom{\Big|_|}$ &
${1\over 18}g_{1\times27}^{} \left( {\cal Q}_{1/2}^V+5{\cal Q}_{3/2}^V\right) - \frac{1}{3}g_{1\times8}^{}{\cal Q}_{1/2}^V$
\\
$\textsl{\texttt O}^{V8,RR}_{uusd}=\textsl{\texttt O}^{V8,RR}_{uusd}\big|_{1\times27}^{} +\textsl{\texttt O}^{V8,RR}_{uusd}\big|_{1\times8}^{} \vphantom{\Big|_|}$~ &
${1\over 54}g_{1\times27}^{} \left( {\cal Q}_{1/2}^V+5{\cal Q}_{3/2}^V\right) + {11\over 36}g_{1\times8}^{}{\cal Q}_{1/2}^V$
\\ \hline
$\textsl{\texttt O}^{V1,LR}_{sduu}=\textsl{\texttt O}^{V1,LR}_{sduu}\big|_{8\times8}^{}
+\textsl{\texttt O}^{V1,LR}_{sduu}\big|_{8\times1}^{} \vphantom{\Big|_|^1}$ &
${1\over6}g_{8\times8}^{} \left( 2{\cal Q}_{1/2}^S+{\cal Q}_{3/2}^S \right)$
\\
$\textsl{\texttt O}^{V8,LR}_{sduu}=\textsl{\texttt O}^{V8,LR}_{sduu}\big|_{8\times8}^{}
+\textsl{\texttt O}^{V8,LR}_{sduu}\big|_{8\times1}^{} \vphantom{\Big|_|}$ &
$-{1\over36} \big(g_{8\times8}^{}-3\tilde g_{8\times8}^{}\big) \left(2{\cal Q}_{1/2}^S+{\cal Q}_{3/2}^S \right)$
\\
$\textsl{\texttt O}^{V1,LR}_{sddd}=\textsl{\texttt O}^{V1,LR}_{sddd}\big|_{8\times8}^{}
 +\textsl{\texttt O}^{V1,LR}_{sddd}\big|_{8\times1}^{} \vphantom{\Big|_|}$ &
$-{1\over12}g_{8\times8}^{} \left( {\cal Q}_{1/2}^S+2{\cal Q}_{3/2}^S \right)$
\\
$\textsl{\texttt O}^{V8,LR}_{sddd}=\textsl{\texttt O}^{V8,LR}_{sddd}\big|_{8\times8}^{}
+\textsl{\texttt O}^{V8,LR}_{sddd}|_{8\times1}^{} \vphantom{\Big|_|}$ &
${1\over72} \big(g_{8\times8}^{}-3\tilde g_{8\times8}^{}\big) \left({\cal Q}_{1/2}^S+2{\cal Q}_{3/2}^S \right)$
\\ \hline
$\textsl{\texttt O}^{V1,LR}_{uusd}=\textsl{\texttt O}^{V1,LR}_{uusd}\big|_{8\times8}^{}
+\textsl{\texttt O}^{V1,LR}_{uusd}\big|_{1\times8}^{} \vphantom{\Big|_|^1}$ &
${1\over6}g_{8\times8}^{} \left(2{\cal Q}_{1/2}^S+{\cal Q}_{3/2}^S \right)$
\\
$\textsl{\texttt O}^{V8,LR}_{uusd}=\textsl{\texttt O}^{V8,LR}_{uusd}\big|_{8\times8}^{}
+\textsl{\texttt O}^{V8,LR}_{uusd}\big|_{1\times8}^{} \vphantom{\Big|_|}$ &
$-{1\over36} \big( g_{8\times8}^{}-3\tilde g_{8\times8}^{} \big) \left(2{\cal Q}_{1/2}^S+{\cal Q}_{3/2}^S \right)$
\\
$\textsl{\texttt O}^{V1,LR}_{ddsd}=\textsl{\texttt O}^{V1,LR}_{ddsd}|_{8\times8}^{}
+\textsl{\texttt O}^{V1,LR}_{ddsd}\big|_{1\times8}^{} \vphantom{\Big|_|}$ &
$-{1\over12}g_{8\times8}^{} \left({\cal Q}_{1/2}^S+2{\cal Q}_{3/2}^S \right)$
\\
$\textsl{\texttt O}^{V8,LR}_{ddsd}=\textsl{\texttt O}^{V8,LR}_{ddsd}|_{8\times8}^{}
+\textsl{\texttt O}^{V8,LR}_{ddsd}\big|_{1\times8}^{} \vphantom{\Big|_|}$ &
${1\over72} \big( g_{8\times8}^{}-3\tilde g_{8\times8}^{}\big) \left({\cal Q}_{1/2}^S+2{\cal Q}_{3/2}^S \right)$
\\ \hline
$\textsl{\texttt O}^{S1,RR}_{uusd}=\textsl{\texttt O}^{S1,RR}_{uusd}\big|_{\bar6\times6} +\textsl{\texttt O}^{S1,RR}_{uusd}\big|_{3\times\bar3} \vphantom{\Big|_|^1}$ &
$-{1\over 24}g_{\bar6\times6}^{} \left(5{\cal Q}_{1/2}^S+4{\cal Q}_{3/2}^S\right) + {1\over8}g_{3\times\bar3}{\cal Q}_{1/2}^S$
\\
$\textsl{\texttt O}^{S8,RR}_{uusd}=\textsl{\texttt O}^{S8,RR}_{uusd}\big|_{\bar6\times6} +\textsl{\texttt O}^{S8,RR}_{uusd}\big|_{3\times\bar3} \vphantom{\Big|_|}$ &
${1\over 144} \big(g_{\bar6\times6} -3\tilde g_{\bar6\times6}\big) \left( 5{\cal Q}_{1/2}^S+4{\cal Q}_{3/2}^S \right) - {1\over 48} \big( g_{3\times\bar3}-3\tilde g_{3\times\bar3} \big) {\cal Q}_{1/2}^S$
\\
$\textsl{\texttt O}^{S1,RR}_{udsu}=\textsl{\texttt O}^{S1,RR}_{udsu}\big|_{\bar6\times6} + \textsl{\texttt O}^{S1,RR}_{udsu}\big|_{3\times\bar3} \vphantom{\Big|_|}$ &
$-{1\over 24}g_{\bar 6\times 6} \left(5{\cal Q}_{1/2}^S+4{\cal Q}_{3/2}^S\right) - \frac{1}{8} g_{3\times \bar3}^{} {\cal Q}_{1/2}^S$ ~
\\
$\textsl{\texttt O}^{S8,RR}_{udsu}=\textsl{\texttt O}^{S8,RR}_{udsu}\big|_{\bar6\times6} +\textsl{\texttt O}^{S8,RR}_{udsu}\big|_{3\times\bar3} \vphantom{\Big|_|}$ &
${1\over 144} \big(g_{\bar6\times6} -3\tilde g_{\bar6\times6}\big) \left(5{\cal Q}_{1/2}^S+4{\cal Q}_{3/2}^S\right) + {1\over 48} \big(g_{3\times\bar3} -3\tilde g_{3\times\bar3}\big) {\cal Q}_{1/2}^S$
\\ \hline
$\textsl{\texttt O}^{S1,LL}_{uusd}=\textsl{\texttt O}^{S1,LL}_{uusd}\big|_{6\times\bar6} +\textsl{\texttt O}^{S1,LL}_{uusd}\big|_{\bar3\times3} \vphantom{\Big|_|^1}$ &
$-{1\over 24}g_{6\times \bar6} \left(5{\cal Q}_{1/2}^S+4{\cal Q}_{3/2}^S\right) + \frac{1}{8} g_{\bar 3\times3} {\cal Q}_{1/2}^S$
\\
$\textsl{\texttt O}^{S8,LL}_{uusd}=\textsl{\texttt O}^{S8,LL}_{uusd}\big|_{6\times\bar6}^{}+\textsl{\texttt O}^{S8,LL}_{uusd}\big|_{\bar3\times3} \vphantom{\Big|_|}$ &
${1\over 144} \big(g_{6\times \bar 6} -3\tilde g_{6\times\bar6}\big) \left(5{\cal Q}_{1/2}^S+4{\cal Q}_{3/2}^S\right) + {1\over 48} \big( g_{\bar 3\times 3} -3\tilde g_{\bar3\times3} \big) {\cal Q}_{1/2}^S$~
\\
$\textsl{\texttt O}^{S1,LL}_{udsu}=\textsl{\texttt O}^{S1,LL}_{udsu}\big|_{6\times\bar6}^{}+\textsl{\texttt O}^{S1,LL}_{udsu}\big|_{\bar3\times3} \vphantom{\Big|_|}$ &
$-{1\over 24} g_{6\times\bar6} \left(5{\cal Q}_{1/2}^S+4{\cal Q}_{3/2}^S\right) - \frac{1}{8}  g_{\bar 3\times3} {\cal Q}_{1/2}^S$
\\
$\textsl{\texttt O}^{S8,LL}_{udsu}=\textsl{\texttt O}^{S8,LL}_{udsu}\big|_{6\times\bar6}^{}+\textsl{\texttt O}^{S8,LL}_{udsu}\big|_{\bar3\times3} \vphantom{\Big|_|}$ &
${1\over 144} \big(g_{6\times\bar 6} -3\tilde g_{6\times\bar6}\big) \left(5{\cal Q}_{1/2}^S+4{\cal Q}_{3/2}^S \right) + {1\over 48} \big(g_{\bar3\times3}-3\tilde g_{\bar3\times3}\big) {\cal Q}_{1/2}^S$
\\ \hline
\end{tabular}
\caption{The chiral representations and realizations of the LEFT four-quark operators contributing to $K\to\pi S(S)$. In the second column, ${\cal Q}_{1/2}^{V,S}$ and ${\cal Q}_{3/2}^{V,S}$ are the mesonic operators defined in eqs.\,\,(\ref{OS}) and (\ref{OV}) and correspond, respectively, to $\Delta I=1/2$ and 3/2 transitions. For $\textsl{\texttt O}^{V1,LR}_{q_1q_2q_3q_4}$ and $\textsl{\texttt O}^{V8,LR}_{q_1q_2q_3q_4}$ in rows 7-14, the contributions of the 8$\times$1 and 1$\times$8 terms are chirally subleading compared to their 8$\times$8 counterparts and therefore dropped from the second column.}
\label{left}
\end{table}

Similarly to what was done in section\,\,\ref{k2pnn}, to examine the impact of each of the dim-6 LEFT operators in table\,\,\ref{smeftleft}, we begin by decomposing their four-quark combinations in terms of the irreducible representations of the chiral group, SU(3)$_L\times$SU(3)$_R$.
Subsequently, for each of the irreps we derive the chiral realization, as prescribed in subsection\,\,\ref{bosonization}, and complement it with a low-energy constant.
Finally, the resulting meson operators are expressed as combinations of their isospin components.
The first column of table\,\,\ref{left} lists the LEFT operators in table\,\,\ref{smeftleft} according to their irreps.
In the second column we collect the mesonic operators pertinent to the $K\to\pi$ processes. The ones corresponding to the four-quark operators with purely left-handed or right-handed quarks are written in terms of the $\Delta I=1/2$ and 3/2 combinations
\begin{align} \label{OV}
{\cal Q}_{1/2}^V & \,=\, F_0^2\left(\partial_\mu K^+\partial^\mu\pi^--{1\over \sqrt{2}}\partial_\mu K^0\partial^\mu\pi^0\right) ,
\nonumber \\
{\cal Q}_{3/2}^V & \,=\, F_0^2\left(\partial_\mu K^+\partial^\mu\pi^-+\sqrt{2}\partial_\mu K^0\partial^\mu\pi^0\right) ,
\end{align}
respectively.
The other entries in the second column involve ${\cal Q}_{1/2}^S$ and ${\cal Q}_{3/2}^S$ which were already defined in eq.\,(\ref{OS}).
More details on the bosonization of the irreps are relegated to appendix\,\,\ref{Appendix:3}.
In this table, we also see that there are more LECs than in table\,\,\ref{tab1}. For $g_{1\times8}^{}$ and $g_{1\times 27}^{}$, which are dimensionless, we adopt
\begin{align}
g_{8\times 1}^{} & \,=\, 3.65 \,, & g_{27\times 1}^{} & \,=\, 0.303\,,~~~ ~~~~
\end{align}
from ref.\,\cite{Cirigliano:2003gt}, whereas $g_{8\times8}^{}$, $\tilde g_{8\times8}^{}$,  $g_{6\times\bar6}^{}$, and $\tilde g_{6\times\bar6}^{}$ are already given in eq.\,(\ref{LEC1}).
Moreover, the parity invariance of the QCD suggests that we can set
\begin{align}
g_{1\times8}^{} & \,=\, g_{8\times 1}^{} \,, &
g_{1\times 27}^{} & \,=\, g_{27\times 1}^{} \,,
\nonumber \\
g_{\bar6\times6}^{} & \,=\, g_{6\times\bar6}^{} \,, &
g_{\bar3\times3}^{} & \,=\, g_{3\times\bar 3}^{} \,, ~~~ ~~~~
\end{align}
and assume analogous relations for the corresponding LECs with a tilde.
For the value of $g_{3\times\bar3}$, there is no estimation yet in literature, and so one can resort to the vacuum saturation approximation (VSA) which yields $g_{3\times\bar3}=g_{6\times\bar6}=B^2_0\simeq4\rm\;GeV^2$ with $B_0 = m^2_\pi/(m_u + m_d) = m^2_K/(m_u + m_s)$ and quark masses at a renormalization scale of 1\,GeV.
Evidently the VSA number for $g_{3\times\bar3}$ is not too far from $g_{6\times \bar6}=3.2\rm\;GeV^2$ in eq.\,(\ref{LEC1}).
Additionally, one can implement simple scaling to estimate $\tilde g_{3\times\bar3}^{}=g_{3\times\bar3}^{}\tilde g_{6\times\bar6}^{}/g_{6\times\bar6}^{}\simeq1.4\rm\;GeV^2$.

It is worth pointing out that, unlike those in table\,\,\ref{tab1}, the operators in table \ref{left} individually either respect the GN inequality or are dominated by portions which do, partly due to the LEC values employed above.
The first two sections of the table contain operators in the latter category as they have ${\cal Q}_{1/2}^V$ parts with $g_{8\times1}^{}\simeq12g_{27\times1}^{}$.
The operators in the remaining sections of the table belong to the first category, such as $\textsl{\texttt O}^{V1,LR}_{sduu}$, which by itself does not affect $K_L\to\pi^0$ transitions due to its chiral realization being proportional to \,$2{\cal Q}_{1/2}^S+{\cal Q}_{3/2}^S=3F_0^2K^+\pi^-$,\, and $\textsl{\texttt O}^{V1,LR}_{sddd}$, which generates ${\cal Q}_{1/2}^S+2{\cal Q}_{3/2}^S\propto K^+\pi^-+K^0\pi^0/\sqrt2$\, whose two terms have the same relative size as their counterparts in ${\cal Q}_{1/2}^S$, albeit with the opposite relative sign.
Nevertheless, there are countless combinations of the various operators which could bring about the breaking of the GN bound.
Simple instances include
\begin{align}
\textsl{\texttt O}^{V,LL}_{ddsd} - \tfrac{6}{11} \textsl{\texttt O}^{V8,LL}_{uusd} & \Rightarrow\,
\tfrac{10}{99}\, g_{27\times1}^{} \Big( {\cal Q}_{1/2}^V - \tfrac{13}{4} {\cal Q}_{3/2}^V \Big)
\,=\, \tfrac{-5}{22}\, g_{27\times1}^{} F_0^2 \Big( \partial_\mu K^+\partial^\mu\pi^-
+ \tfrac{5\sqrt2}{3}\, \partial_\mu K^0\partial^\mu\pi^0 \Big) \,,
\nonumber \\
\textsl{\texttt O}^{V1,LR}_{sduu} + 3 \textsl{\texttt O}^{V1,LR}_{sddd} & \Rightarrow\,
\tfrac{1}{12}\, g_{8\times8}^{} \big( {\cal Q}_{1/2}^S-4 {\cal Q}_{3/2}^S \big) \,=\,
-\tfrac{1}{4}\, g_{8\times8}^{} F_0^2 \Big( K^+\pi^- + \tfrac{3}{\sqrt2} K^0\pi^0 \Big) \,.
\end{align}
We conclude that judicious choices for the coefficients of the operators would need to be made in order to evade the bound in a significant manner.\footnote{This conclusion is in line with what has been argued qualitatively in \cite{Ziegler:2020ize}, namely that for heavy mediators four-quark operators, especially those of the types in the first two sections of table \ref{left}, might not be able to cause \,$K\to\pi S$\, decays to violate the GN bound by more than a factor of a few.
As our examples demonstrate, preference for the \,$\Delta I=3/2$\, components of the operators is required to achieve substantial violations.}
We will encounter more examples later on.

\subsection{\bf Numerical analysis}

Summing the mesonic operators in table\,\,\ref{left} multiplied by their respective Wilson coefficients leads to the effective Lagrangian ${\cal L}_{K\pi S}$ responsible for $K\to\pi S$. We can express it as
\begin{align}
{\cal L}_{K\pi S} & \,=\, F_0 \bigg(a_1 K^+\pi^- -{b_1\over \sqrt{2}} K^0\pi^0 \bigg) S+{1\over F_0}\bigg(a_2 \partial_\mu K^+\partial^\mu \pi^--{b_2\over \sqrt{2}}\partial_\mu K^0\partial^\mu \pi^0\bigg) S \,+\, {\rm H.c.}
\nonumber
\\ & \,\supset\, F_0 \big[ a_1 K^+\pi^--({\rm Re}\,b_1) K_L\pi^0 \big] S + \frac{1}{F_0} \big[ a_2 \partial_\mu K^+\partial^\mu \pi^--({\rm Re}\,b_2)\partial_\mu  K_L\partial^\mu \pi^0 \big] S \,,
\end{align}
where $a_{1,2}$ and $b_{1,2}$ are dimensionless constants comprising linear combinations of the Wilson coefficients ${\tt C}$s, namely
\begin{align} \nonumber
a_1 \,=\, \frac{1}{24} F_0\, & \bigg[ 6 \Big( 2{\tt C}_{sduu}^{V1,LR}- {\tt C}_{sddd}^{V1,LR}+2{\tt C}_{uusd}^{V1,LR} -{\tt C}_{ddsd}^{V1,LR} \Big) g_{8\times8}^{}
\\ \nonumber
& - \Big(2{\tt C}_{sduu}^{V8,LR}- {\tt C}_{sddd}^{V8,LR}+2{\tt C}_{uusd}^{V8,LR} -{\tt C}_{ddsd}^{V8,LR} \Big) \big( g_{8\times8}^{}-3\tilde g_{8\times8}^{} \big)
\\\nonumber
& - 9\left({\tt C}_{uusd}^{S1,LL}+{\tt C}_{udsu}^{S1,LL}+{\tt C}_{uusd}^{S1,RR}+{\tt C}_{udsu}^{S1,RR}\right) g_{6\times\bar6}^{}
\\\nonumber
& + {3\over 2}\left({\tt C}_{uusd}^{S8,LL}+{\tt C}_{udsu}^{S8,LL}+{\tt C}_{uusd}^{S8,RR}+{\tt C}_{udsu}^{S8,RR}\right) \left(g_{6\times\bar6}^{}-3\tilde g_{6\times\bar6}^{}\right)
\\\nonumber
& + 3\left({\tt C}_{uusd}^{S1,LL}-{\tt C}_{udsu}^{S1,LL}+{\tt C}_{uusd}^{S1,RR}-{\tt C}_{udsu}^{S1,RR}\right) g_{3\times \bar 3}
\\
& - {1\over 2}\left({\tt C}_{uusd}^{S8,LL}-{\tt C}_{udsu}^{S8,LL}+{\tt C}_{uusd}^{S8,RR}-{\tt C}_{udsu}^{S8,RR}\right) \left(g_{3\times\bar3}^{} - 3 \tilde g_{3\times\bar3}^{}\right) \bigg]_{\Lambda_\chi} \,, & \hspace{3em} \label{a_1}
\end{align}
\begin{align} \nonumber
b_1 \,=\, \frac{1}{24} F_0\, & \bigg[ 6\left( {\tt C}_{sddd}^{V1,LR} + {\tt C}_{ddsd}^{V1,LR} \right)g_{8\times8}^{}
- \left( {\tt C}_{sddd}^{V8,LR}+{\tt C}_{ddsd}^{V8,LR} \right) \left( g_{8\times8}^{}-3\tilde g_{8\times8}^{} \right)
\\\nonumber
& + 3\left({\tt C}_{uusd}^{S1,LL}+{\tt C}_{udsu}^{S1,LL}+{\tt C}_{uusd}^{S1,RR}+{\tt C}_{udsu}^{S1,RR}\right)g_{6\times\bar6}^{}
\\\nonumber
& - {1\over 2}\left({\tt C}_{uusd}^{S8,LL}+{\tt C}_{udsu}^{S8,LL}+{\tt C}_{uusd}^{S8,RR}+{\tt C}_{udsu}^{S8,RR}\right)\left(g_{6\times\bar6}^{}-3\tilde g_{6\times\bar6}^{}\right)
\\\nonumber
& + 3\left({\tt C}_{uusd}^{S1,LL}-{\tt C}_{udsu}^{S1,LL}+{\tt C}_{uusd}^{S1,RR}-{\tt C}_{udsu}^{S1,RR}\right)g_{3\times \bar 3}
\\
& - {1\over 2}\left({\tt C}_{uusd}^{S8,LL}-{\tt C}_{udsu}^{S8,LL}+{\tt C}_{uusd}^{S8,RR}-{\tt C}_{udsu}^{S8,RR}\right)g_{3\times \bar 3} \bigg]_{\Lambda_\chi} \,, & \hspace{3em} \label{b_1}
\end{align}
\begin{align} \nonumber
a_2 \,=\, \frac{1}{36} F_0^3\, & \Big[ 6\left( {\tt C}_{ddsd}^{V,LL}+{\tt C}_{ddsd}^{V,RR}-2 {\tt C}_{uusd}^{V1,LL}-2{\tt C}_{uusd}^{V1,RR}\right)\left(g_{8\times1}^{}-g_{27\times1}^{}\right)
\\
& + \left({\tt C}_{uusd}^{V8,LL}+{\tt C}_{uusd}^{V8,RR}\right)\left(11g_{8\times1}^{}+4g_{27\times1}^{}\right) \Big]_{\Lambda_\chi} \,, & \hspace{7em} \label{a_2}
\end{align}
\begin{align} \nonumber
b_2 \,=\, \frac{1}{36} F_0^3\, & \Big[ 6\left( {\tt C}_{ddsd}^{V,LL}+{\tt C}_{ddsd}^{V,RR}\right)\left(g_{8\times1}^{}+4g_{27\times1}^{}\right)-6\left( {\tt C}_{uusd}^{V1,LL}+{\tt C}_{uusd}^{V1,RR}\right)\left(2g_{8\times1}^{}+3g_{27\times1}^{}\right)
\\
& + \left({\tt C}_{uusd}^{V8,LL}+{\tt C}_{uusd}^{V8,RR}\right)\left(11g_{8\times1}^{}-6g_{27\times1}^{}\right) \Big]_{\Lambda_\chi} \,, & \label{b_2}
\end{align}
the subscript $\Lambda_\chi$ indicating that the ${\tt C}$s on the right-hand sides are evaluated at $\mu=\Lambda_\chi$.
These coefficients scale as $\Lambda_{\textsc{np}}^{-3}$.

For $K\to\pi SS$, the interaction Lagrangian ${\cal L}_{K\pi SS}$ has an expression similar to $S{\cal L}_{K\pi S}$, namely
\begin{align}
{\cal L}_{K\pi SS} & \,= \Bigg( \hat a_1 K^+\pi^- - \frac{\hat b_1}{\sqrt2}\, K^0\pi^0  \Bigg) S^2 + \Bigg( \hat a_2\, \partial_\mu K^+\partial^\mu \pi^- - \frac{\hat b_2}{\sqrt2}\, \partial_\mu K^0\partial^\mu \pi^0 \Bigg) \frac{S^2}{F_0^2} \,+\, {\rm H.c.}\;.
\end{align}
The dimensionless parameters $\hat a_{1,2}$ and $\hat b_{1,2}$ are the same in form as $a_{1,2} F_0$ and $b_{1,2}F_0$, respectively, but with the Wilson coefficients now denoted by $\hat{\tt C}$s, which scale as $\Lambda_{\textsc{np}}^{-4}$ because the underlying quark-level operators are of dimension eight.

\subsubsection{\boldmath$K\to\pi S$} 
From ${\cal L}_{K\pi S}$, we obtain the amplitudes for $K\to \pi S$ to be
\begin{align}
{\cal A}_{K^+\to\pi^+S} & \,=\, a_1 F_0+ a_2\, {m_{K^+}^2+m_{\pi^+}^2-m_S^2\over 2 F_0} \,,
\nonumber \\
{\cal A}_{K_L\to\pi^0S} & \,=\, -{\rm Re}\,b_1\, F_0 - {\rm Re}\,b_2\, {m_{K^0}^2+m_{\pi^0}^2-m_S^2\over 2 F_0} \,,
\end{align}
and hence the branching fractions
\begin{align}
{\cal B}(K^+\to\pi^+S) & \,=\, \tau_{K^+}\, \frac{\sqrt{(m_{K^+}^2-m_{\pi^+}^2)^2 -(2m_{K^+}^2+2m_{\pi^+}^2-m_S^2)m_S^2}}{16\pi m_{K^+}^3}
\nonumber \\ & ~~~\times\, \Bigg|a_1+a_2\,\frac{m_{K^+}^2+m_{\pi^+}^2-m_S^2}{2F_0^2}\Bigg|^2 F_0^2\,,
\\
{\cal B}(K_L\to\pi^0S) & \,=\, \tau_{K_L}\, \frac{\sqrt{(m_{K^0}^2-m_{\pi^0}^2)^2 -(2m_{K^0}^2+2m_{\pi^0}^2-m_S^2)m_S^2}}{16\pi m_{K^0}^3}^{\vphantom{|^|}}
\nonumber \\ & ~~~\times\, \Bigg|{\rm Re}\,b_1 + {\rm Re}\,b_2\, \frac{m_{K^0}^2+m_{\pi^0}^2-m_S^2}{2 F_0^2} \Bigg|^2 F_0^2 \,.
\end{align}
To account for the KOTO anomaly, one could consider various possibilities.
For illustration, we look at a scenario in which the only contributing operators are those with purely right-handed quarks: $\textsl{\texttt O}_{ddsd}^{V,RR}$ and $\textsl{\texttt O}_{uusd}^{(V1,V8),RR}$.
This implies that $a_1=b_1=0$ and
\begin{align}\nonumber
a_2 & \,=\, {F_0^3\over6} \bigg[ \bigg({\tt C}_{ddsd}^{V,RR}-2{\tt C}_{uusd}^{V1,RR}+{11\over6}{\tt C}_{uusd}^{V8,RR}\bigg) g_{8\times1}^{} - \bigg({\tt C}_{ddsd}^{V,RR}-2{\tt C}_{uusd}^{V1,RR}-{2\over 3}{\tt C}_{uusd}^{V8,RR}\bigg) g_{27\times1}^{} \bigg]_{\Lambda_\chi} ,
\\ \label{a2b2}
b_2 & \,=\, {F_0^3\over6} \bigg[ \bigg({\tt C}_{ddsd}^{V,RR}-2{\tt C}_{uusd}^{V1,RR}+{11\over6}{\tt C}_{uusd}^{V8,RR}\bigg) g_{8\times1}^{}
+ \bigg(4{\tt C}_{ddsd}^{V,RR}-3{\tt C}_{uusd}^{V1,RR}-{\tt C}_{uusd}^{V8,RR}\bigg) g_{27\times1}^{}\bigg]_{\Lambda_\chi} .
\end{align}
Moreover, we select \,${\tt C}_{ddsd}^{V,RR}=2{\tt C}_{uusd}^{V1,RR}-(11/6){\tt C}_{uusd}^{V8,RR}$\, to make the $g_{8\times1}^{}$ terms above vanish, changing eq.\,(\ref{a2b2}) to
\begin{align} \label{a2'b2'}
a_2 & \,=\,  {5F_0^3\over12}g_{27\times1}^{}{\tt C}_{uusd}^{V8,RR}(\Lambda_\chi)\;, &
b_2 & \,=\, {25F_0^3\over18}g_{27\times1}^{} \left( \frac{3}{5}\, {\tt C}_{uusd}^{V1,RR}(\Lambda_\chi) - {\tt C}_{uusd}^{V8,RR}(\Lambda_\chi) \right) ,
\end{align}
with which, for $m_S=0$, we arrive at
\begin{align} \label{rBNP}
\tilde r_{\cal B}^{\rm NP} &= \frac{{\cal B}(K_L\to\pi^0S)}{{\cal B}(K^+\to\pi^+S)} = 4.13\, {(17.6\,{\rm Re}\, b_2)^2 \over \left|17.4\,a_2\right|^2 } = 47 \left| \frac{{\rm Re} \Big[ {\tt C}_{uusd}^{V8,RR}(\Lambda_\chi) - 0.6\, {\tt C}_{uusd}^{V1,RR}(\Lambda_\chi) \Big]^{\vphantom{|}} }{{\tt C}_{uusd}^{V8,RR}(\Lambda_\chi)} \right|^2 .
\end{align}
It is worth remarking that the potential enlargement of $\tilde r_{\cal B}^{\rm NP}$ in this equation can be expected from the fact that it arises from the quark operators $\textsl{\texttt O}^{V,RR}_{ddsd}$ and $\textsl{\texttt O}^{(V1,V8),RR}_{uusd}$ which, as rows 4-6 in table\,\,\ref{left} show, in the absence of the $g_{8\times1}^{}$ portions, generate the combinations $2{\cal Q}_{1/2}^V-5{\cal Q}_{3/2}^V$ and ${\cal Q}_{1/2}^V+5{\cal Q}_{3/2}^V$ of the mesonic operators defined in eq.\,(\ref{OV}) and hence all contain significant $\Delta I=3/2$ components. Clearly, a much amplified $\tilde r_{\cal B}^{\rm NP}$ can be easily realized with some more tuning of the parameters in eq.\,(\ref{rBNP}).

\begin{figure}[t]
\includegraphics[height=6cm,width=8.6cm]{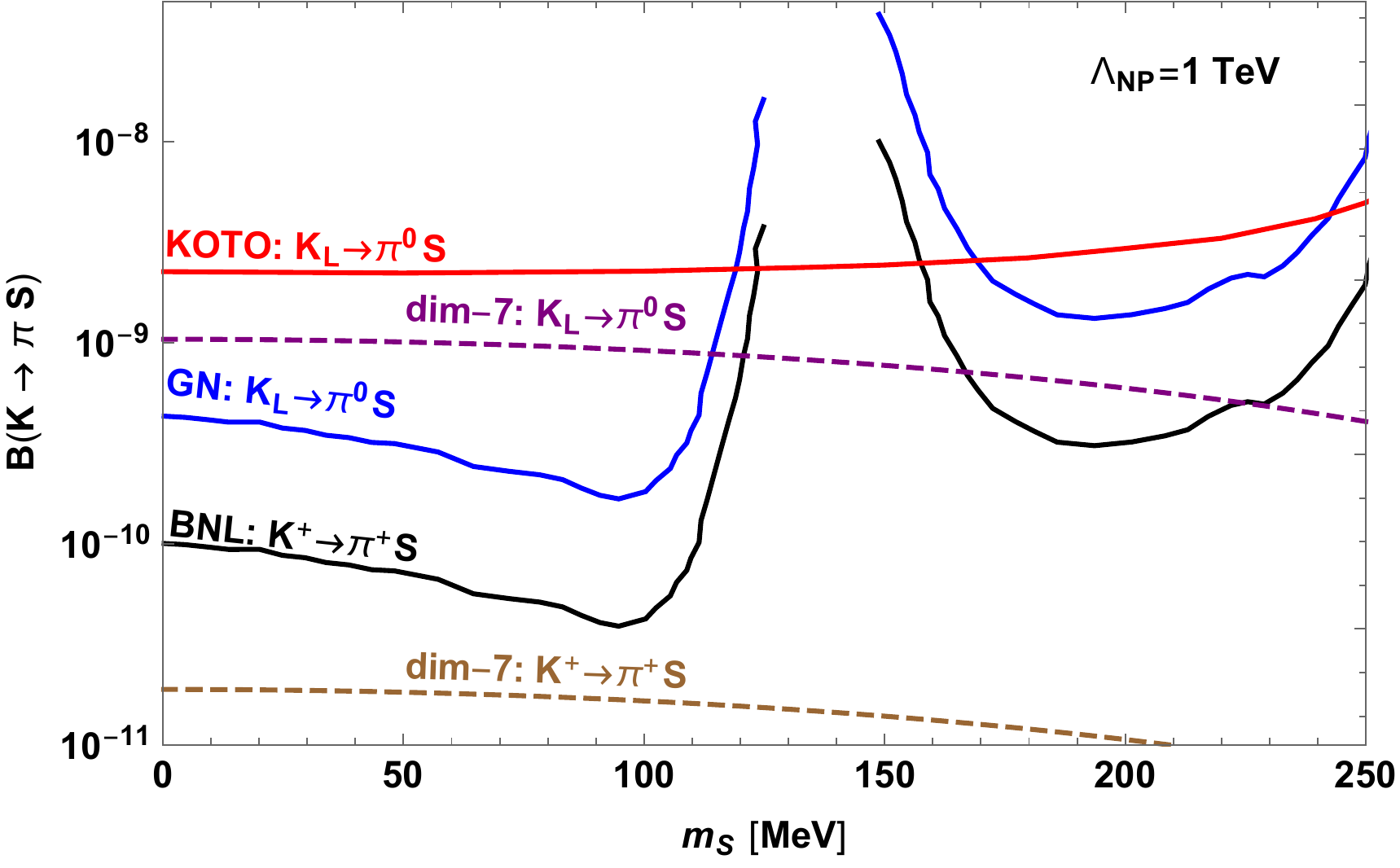}
\includegraphics[height=6cm,width=8.6cm]{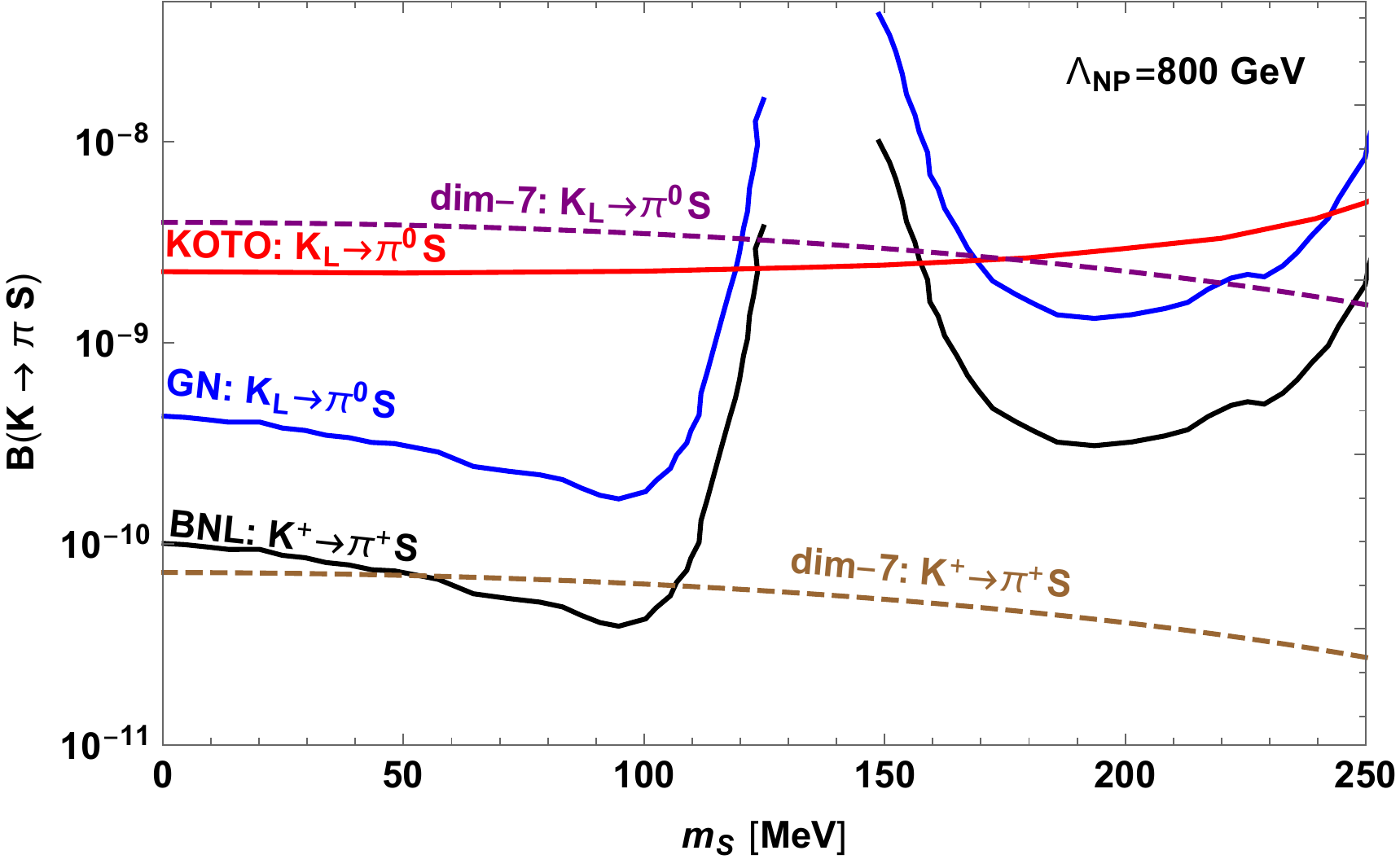} \vspace{-1ex}
\caption{The branching fractions of $K\to\pi S$ versus the $S$ mass $m_S$ induced by the dim-7 operators with NP scales $\Lambda_{\textsc{np}}=1$\,\,TeV (left panel) and 800\,\,GeV (right  panel), compared to the experimental upper limits from KOTO 2015~\cite{Ahn:2018mvc} and BNL~\cite{Artamonov:2009sz} along with the standard GN constraint on $K_L\to \pi^0S$ from the BNL result. The blue and black curves are related by ${\cal B}(K_L\to \pi^0S)_{\rm GN}=4.3\,{\cal B}(K^+\to\pi^+S)_{\rm BNL}$.}
\label{fig1}
\end{figure}

To be more precise in our numerical treatment, we again must take into account the QCD RG running of the coefficients from the EW scale, which we choose to be the $W$-boson mass $m_W$ as before, down to the chiral-symmetry breaking scale $\Lambda_\chi$.
The pertinent one-loop RG equations are available in ref.\,\,\cite{Jenkins:2017dyc}, from which we collect the formulas in appendix\,\,\ref{rge}. We use them to get
\begin{align}\nonumber
{\tt C}_{uusd}^{V1,RR}(\Lambda_\chi) & \,=\, 1.07\, {\tt C}_{uusd}^{V1,RR}(m_W) - 0.19\, {\tt C}_{uusd}^{V8,RR}(m_W) \,,
\\
{\tt C}_{uusd}^{V8,RR}(\Lambda_\chi) & \,=\, 1.31\, {\tt C}_{uusd}^{V8,RR}(m_W) - 0.86\, {\tt C}_{uusd}^{V1,RR}(m_W) - 0.16\, {\tt C}_{ddsd}^{V,RR}(m_W) \,,
\end{align}
which enter eq.\,(\ref{a2'b2'}) and depend on other coefficients. To simplify things further, we can pick ${\tt C}_{uusd}^{V1,RR}(m_W)={\tt C}_{ddsd}^{V,RR}(m_W)=0$\, and \,${\tt C}_{uusd}^{V8,RR}(m_W)=\Lambda_{\textsc{np}}^{-3}$,\, which lead to \,$\tilde r_{\cal B}^{\rm NP}=51$\, in eq.\,(\ref{rBNP}).
This exceeds the maximum \,$r_{\cal B}^{\rm GN}=4.3$\, of the GN bound by more than 10 times.
For this example, in figure\,\,\ref{fig1} we depict ${\cal B}(K\to\pi S)$ as functions of the $S$ mass $m_S$ with \,$\Lambda_{\textsc{np}}=1$\,\,TeV (left panel) and 800\,\,GeV (right panel).
In the figure, we also exhibit the upper limits on ${\cal B}(K_L\to\pi^0S)$ and ${\cal B}(K^+\to \pi^+S)$ at 90\% CL from KOTO 2015~\cite{Ahn:2018mvc} and BNL~\cite{Artamonov:2009sz}, respectively, along with the empirical GN constraint based on the BNL result: ${\cal B}(K_L\to \pi^0S)_{\rm GN}<4.3\,{\cal B}(K^+\to\pi^+S)_{\rm BNL}$,\, which is reflected by the blue and black curves in each graph.
If \,$\Lambda_{\textsc{np}}=1$\,\,TeV,\, the left panel reveals that the GN inequality is not respected in the \,$m_S \,\raisebox{1pt}{\footnotesize$\lesssim$}\, 110$\,\,MeV\, region, while the current experimental limits are satisfied.
If $\Lambda_{\textsc{np}}$ is smaller, it is possible to break the GN bound with higher $m_S$ values, such as \,$170{\rm\;MeV} \,\raisebox{1pt}{\footnotesize$\lesssim$}\, m_S \,\raisebox{1pt}{\footnotesize$\lesssim$}\, 240$\,\,MeV\, in the right panel for \,$\Lambda_{\textsc{np}}=800$\,\,GeV.\,
We conclude that it can be violated by dim-7 EFT operators with a NP scale $\Lambda_{\textsc{np}}=\cal O$(1\,\,TeV).

\subsubsection{\boldmath$K\to\pi SS$}

\begin{figure}[h] \bigskip
\includegraphics[width=11cm]{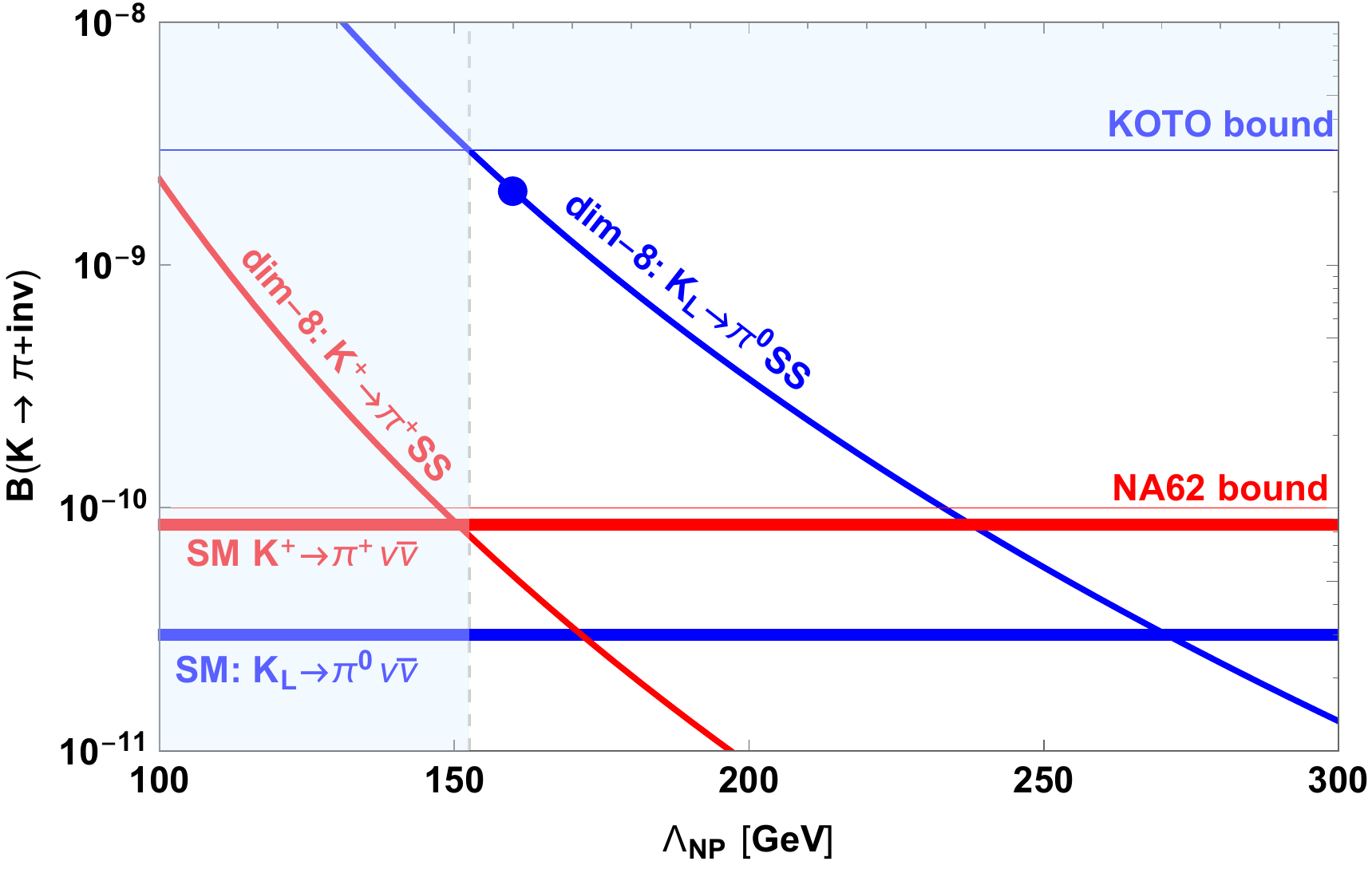} \vspace{-1ex}
\caption{The branching fractions of $K\to\pi SS$ induced by the dim-8 operators as functions of the NP scale~$\Lambda_{\textsc{np}}$ for $m_S=0$, as described in the text.
Also displayed are the corresponding SM predictions for  $K\to\pi\nu\bar\nu$ (red and blue horizontal bands) and upper limits from KOTO \cite{Ahn:2018mvc} and NA62 \cite{na62} (blue and red horizontal thin lines). The light-blue region is excluded by the KOTO bound. The blue dot corresponds to KOTO's three events.\label{KtopiSS}}   \bigskip
\end{figure}

For the three-body decays $K\to\pi SS$, the amplitudes are
\begin{align} \label{Ak2pss}
{\cal A}_{K^+\to\pi^+SS} & \,=\,  2\hat a_1 + \hat a_2\, {m_{K^+}^2+m_{\pi^+}^2-\hat s\over F_0^2}\,, &
\nonumber \\
{\cal A}_{K_L\to\pi^0SS} & \,=\, -2\,{\rm Re}\,\hat b_1 - {\rm Re}\,\hat b_2\, {m_{K^0}^2+m_{\pi^0}^2-\hat s\over F_0^2} \,,
\end{align}
where $\hat s$ designates the invariant mass squared of the $SS$ pair.
These bring about another modified definition of $r_{\cal B}$,
\begin{align} \label{rBSS}
r_{\cal B} & \,=\, \frac{{\cal B}(K_L\to\pi^0\mbox{+}E_{\rm miss})}
{{\cal B}(K^+\to\pi^+\mbox{+}E_{\rm miss})}
\,=\, \frac{{\cal B}(K_L\to\pi^0\nu\bar\nu)_{\rm SM}^{} + {\cal B}(K_L\to\pi^0SS)}
{{\cal B}(K^+\to\pi^+\nu\bar\nu)_{\rm SM}^{} + {\cal B}(K^+\to\pi^+SS)} \,. ~~~
\end{align}
As before, we have many options regarding the parameters which can yield a violation of the GN bound.
This time we entertain the possibility that the operators with purely left-handed or right-handed quarks do not contribute, implying that \,$\hat a_2=\hat b_2=0$\, and so the branching fractions become,  for $m_S=0$,
\begin{align}\nonumber
{\cal B}(K^+\to\pi^+SS) & \,=\, {\tau_{K^+}\over 2^9\pi^3m_{K^+}^3}\int\Pi_3|{\cal M}_{K^+\to\pi^+2S}|^2
\,=\, {\tau_{K^+}|\hat a_1|^2\over 2^7\pi^3m_{K^+}^3}\int\Pi_3
\,=\, 7\times 10^{11}|\hat a_1|^2\;,
\\ \label{exk2pss}
{\cal B}(K_L\to\pi^0SS) & \,=\, {\tau_{K_L}\over 2^9\pi^3m_{K^0}^3}\int\Pi_3|{\cal M}_{K_L\to\pi^02S}|^2
\,=\, {\tau_{K_L}({\rm Re}\,\hat  b_1)^2 \over 2^7\pi^3m_{K^0}^3}\int\Pi_3
\,=\, 3\times 10^{12}({\rm Re}\,\hat b_1)^2 \,.
\end{align}
Supposing additionally that $\hat a_1$ and $\hat b_1$ come only from the $g_{8\times8}^{}$ terms in eqs.\,\,(\ref{a_1}) and (\ref{b_1}) and setting
\,${\tt C}_{sddd}^{V1,LR}/2={\tt C}_{ddsd}^{V1,LR}={\tt C}_{uusd}^{V1,LR}={\tt C}_{sduu}^{V1,LR}=\Lambda_{\textsc{np}}^{-4}$,\, we derive
\begin{align} \label{ex'k2pss}
\hat a_1 & \,=\, {F_0^2\over 4}\, {g_{8\times 8}^{}\over \Lambda_{\textsc{np}}^4} \,, &
\hat b_1 & \,=\, {F_0^2\over 4}\, {3g_{8\times 8}^{}\over \Lambda_{\textsc{np}}^4}\,. ~~~ ~~~~
\end{align}
In figure\,\,\ref{KtopiSS} we plot the resulting branching fractions of $K\to\pi SS$ as functions of $\Lambda_{\textsc{np}}$.
Evidently, in this particular instance, to boost ${\cal B}(K_L\to\pi^0SS)$ to a level within KOTO's current sensitivity reach would need $\Lambda_{\textsc{np}}$ to be no more than roughly 200\,\,GeV.
As can be inferred from this graph in conjunction with eq.\,(\ref{rBSS}), for $\Lambda_{\textsc{np}}$ above this value the GN bound is no longer violated.

\section{Summary and Conclusions\label{concl}}

Motivated by the recent preliminary observation of three anomalous events of $K_L\to\pi^0\nu\bar\nu$ by the KOTO Collaboration, we study in detail the possibility of having new physics responsible for enhancing the $K\to\pi$+$E_{\rm miss}$ modes over their SM expectations. We explore two types of scenarios:
\begin{itemize}
\item NP above the EW scale represented by quark-neutrino interactions which do not preserve lepton flavor/number.
\item NP above the EW scale with new scalar particles that are sufficiently light to be produced in $K\to\pi$+$E_{\rm miss}$ decays.
\end{itemize}
The NP is described with an effective Lagrangian above the EW scale that respects the gauge symmetries of the SM. In all the cases considered, we specifically look for true violations of the Grossman-Nir bound through four-quark $\Delta I = 3/2$ interactions.

The NP effects are classified according to the mass dimensionality of the necessary operators. To this end, we catalogue all the operators that can give rise to the reaction $K\to\pi\cal X$ with $\cal X$ standing for one or more particles carrying away the missing energy.
As itemized above, we allow $\cal X$ to comprise: a neutrino-antineutrino pair ($\nu\bar\nu$), a pair of neutrinos ($\nu\nu$) or antineutrinos~($\bar\nu\bar\nu$), an invisible light new scalar boson ($S$), and a pair of these scalars ($SS$).
These cases require a~minimal dimensionality of ten, nine, seven, and eight, respectively.
On general grounds, we argue that the scenarios with new scalars (dim-7 or -8 operators) are consistent with sizable boosts in the rate of \,$K_L\to\pi^0$+$E_{\rm miss}$\, for NP scales above the EW scale.

We construct the effective Lagrangian for each of the cases and, after identifying the $\Delta I = 3/2$ components of the operators, we discuss the renormalization group running of the couplings down to a hadronic scale followed by the matching of the operators onto chiral perturbation theory. We present numerical results illustrating the scale of the NP needed to amplify the \,$K_L\to\pi^0$+$E_{\rm miss}$\, rate above the GN bound obtained from the measurements of \,$K^+\to\pi^+$+$E_{\rm miss}$ .

We find that the production of a single light new scalar via $\Delta I = 3/2$ interactions permits enlargements in the \,$K_L\to\pi^0$+$E_{\rm miss}$\, rate that are big enough to appear in the KOTO experiment, and we clarify this with figure\,\,\ref{fig1}. This is achievable with some degree of tuning among the coefficients of the operators.
Our results are attained for stable new scalars, but long-lived ones would also work as they have weaker constraints \cite{Artamonov:2009sz}.

\begin{figure}[t]
\includegraphics[width=11cm]{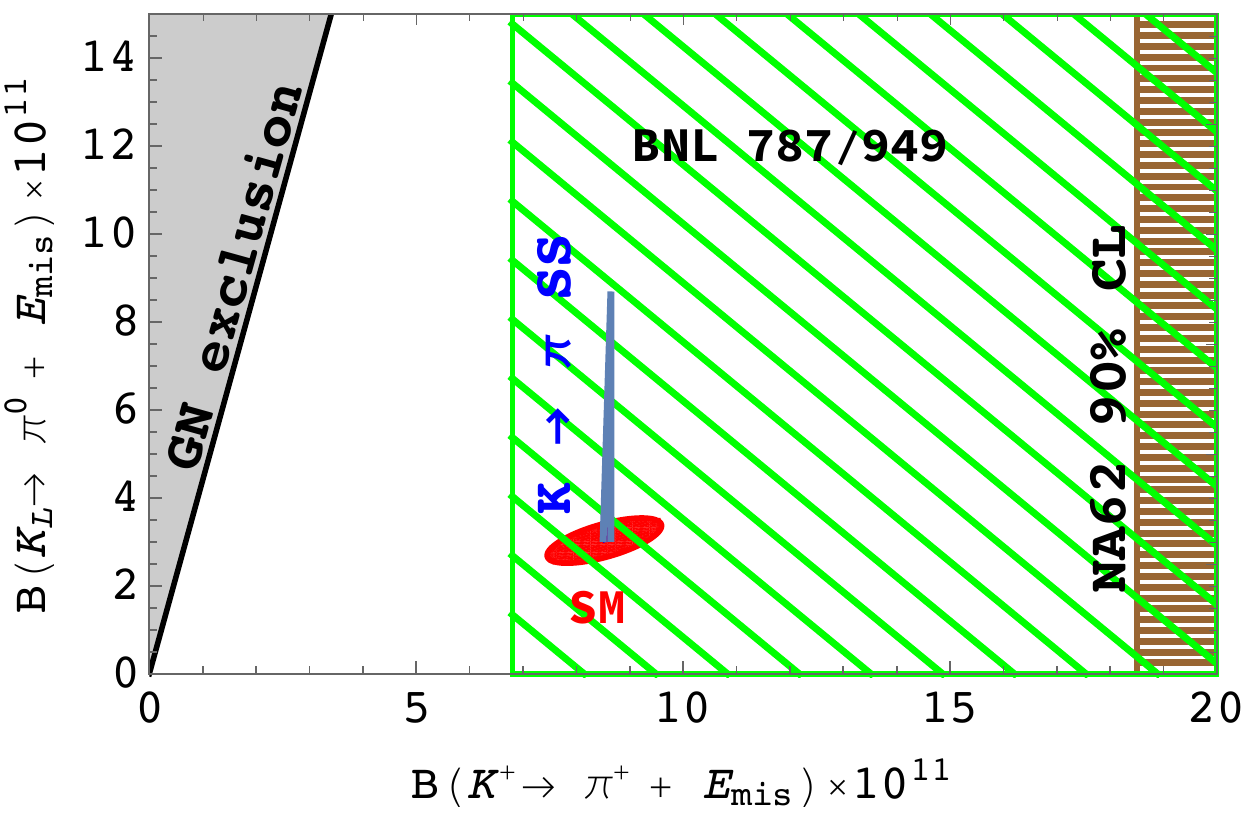} \vspace{-1ex}
\caption{The branching fractions of $K\to\pi$+$E_{\rm miss}$: in red the 90\%-CL SM predictions; in green the $1\sigma$ BNL E787/E949 result; in brown the 90\% NA62 exclusion; in grey the GN bound; and in blue a~region accessible with $K\to \pi SS$ for parameters chosen to enhance mostly the neutral mode with a NP scale~$\Lambda_{\textsc{np}}\geq v$.\label{f:xss}}
\end{figure}

The production of a pair of the new light scalars could have substantial rate gains over the SM but not above the GN bound.
We depict this in figure\,\,\ref{f:xss},\footnote{To draw the blue region, we use again the example in eq.\,(\ref{exk2pss}), with
\,$\hat a_1=F_0^2 g_{8\times 8}^{}/\big(4\Lambda_{\textsc{np}}^4\big)$\, as in eq.\,(\ref{ex'k2pss}),
but now let ${\rm Re}\,\hat b_1$ vary under the condition \,$0\le|{\rm Re}\,\hat b_1|\le3|\hat a_1|$\, and demand \,$\Lambda_{\textsc{np}}\ge v$.
With regard to the operator coefficients in eqs.\,\,(\ref{a_1}) and (\ref{b_1}), one way to accomplish this is to arrange \,$2{\tt C}_{sduu}^{V1,LR}- {\tt C}_{sddd}^{V1,LR}+2{\tt C}_{uusd}^{V1,LR} -{\tt C}_{ddsd}^{V1,LR}=\Lambda_{\textsc{np}}^{-4}$ and~$\,0\le {\tt C}_{sddd}^{V1,LR} + {\tt C}_{ddsd}^{V1,LR}\le3$,\, having taken the others to vanish.} where the blue area illustrates that increases over the SM by factors of a few are possible while keeping \,$\Lambda_{\textsc{np}}\ge v$\, [in contrast, to exceed the bound (explain the KOTO events) would require \,$\Lambda_{\textsc{np}}<200\;$GeV ($\Lambda_{\textsc{np}}\sim160\;$GeV),\, as indicated in figure\,\,\ref{KtopiSS}].
With a different choice of parameters, the charged mode could also be amplified by a similar amount.

The production of two neutrinos, on the other hand, suffers from relatively much greater $\Lambda_{\textsc{np}}$ suppression. The restriction $\Lambda_{\textsc{np}}\geq v$ results in very small rises over the SM, completely within the uncertainty of the SM predictions and thus unobservable.

We conclude that continued improvement of the KOTO upper bound on $K_L\to\pi^0$+$E_{\rm miss}$, even at current levels which are much above the GN bound, provides relevant constraints on possible new physics scenarios.

\acknowledgements
This work was supported in part by the MOST (Grant No. MOST 106-2112-M-002-003-MY3).
This work was also supported in part by the Australian Government through the Australian Research Council.

\appendix

\section{Isospin decomposition of quark parts of dim-9 operators\label{isodec}}

In this appendix, for completeness we write down the decomposition of the quark portion of each of the dim-9 operators
in eq.\,(\ref{opebasis2}) into its $\Delta I=1/2,3/2$ components.
This will allow us to see clearly the difference between them.
Since additionally each operator also causes a definite change $\Delta I_3$ in the third isospin component, we can first group them according to their $\Delta I_3$ values and then express them as linear combinations of their $\Delta I$ terms. Inspecting the operators, we find that ${\texttt O}_1^{usdu}$, $\tilde{\texttt O}{}_1^{usdu}$, ${\texttt O}_3^{ddds}$, and $\tilde{\texttt O}{}_3^{ddds}$ have $\Delta I_3=1/2$, whereas ${\texttt O}_1^{udsu}$, $\tilde{\texttt O}{}_1^{udsu}$, ${\texttt O}_3^{ddsd}$, and $\tilde{\texttt O}{}_3^{ddsd}$ have $\Delta I_3=-1/2$.
Employing the Clebsch-Gordan decomposition rule, we then get the following results:
\begin{itemize}
\item The $\Delta I_3=1/2$ operators:
\begin{align}
{\texttt O}_1^{usdu} &=-{1\over3} {\texttt O}_{1,\Delta I=1/2}^{usdu}+{1 \over 3}{\texttt O}_{1,\Delta I=3/2}^{usdu}\;, &
{\texttt O}_3^{ddds} &=\frac{1}{3}{\texttt O}_{3,\Delta I=1/2}^{ddds}-{1 \over 3}{\texttt O}_{3,\Delta I=3/2}^{ddds} \;,
\end{align}
with their components of definite $\Delta I$ being given by
\begin{align}\nonumber
 {\texttt O}_{1,\Delta I=1/2}^{usdu} & = \Big[ \big(\overline{d_L}\gamma_\mu s_L\big) \big( \overline{u_R}\gamma^\mu u_R\big)
-2 \big(\overline{u_L}\gamma_\mu s_L\big)\big( \overline{d_R}\gamma^\mu u_R\big)
- \big(\overline{d_L}\gamma_\mu s_L\big)\big( \overline{d_R}\gamma^\mu d_R\big) \Big]J \,,
 \\\nonumber
 {\texttt O}_{1,\Delta I=3/2}^{usdu} & = \Big[ \big(\overline{d_L}\gamma_\mu s_L\big) \big( \overline{u_R}\gamma^\mu u_R\big)
+ \big(\overline{u_L}\gamma_\mu s_L\big) \big( \overline{d_R}\gamma^\mu u_R\big)
- \big(\overline{d_L}\gamma_\mu s_L\big) \big( \overline{d_R}\gamma^\mu d_R\big) \Big]J \,,
 \\\nonumber
{\texttt O}_{3,\Delta I=1/2}^{ddds} & = \Big[ \big(\overline{u_R}u_L \big) \big(\overline{d_R}s_L) 
+ \big(\overline{d_R}u_L\big) \big(\overline{u_R}s_L\big) + 2 \big(\overline{d_R}d_L\big) \big(\overline{d_R}s_L\big) \Big]J \,,
\\
{\texttt O}_{3,\Delta I=3/2}^{ddds}&= \Big[ \big(\overline{u_R}u_L\big) \big(\overline{d_R}s_L\big)
+ \big(\overline{d_R}u_L\big) \big(\overline{u_R}s_L\big) - \big(\overline{d_R}d_L\big) \big(\overline{d_R}s_L\big) \Big]J \,,
\end{align}
and similarly $\tilde{\texttt O}{}_1^{usdu}$ and $\tilde{\texttt O}{}_3^{ddds}$.
\item The $\Delta I_3=-1/2$ operators:
\begin{align}
{\texttt O}_1^{udsu}&=-{1 \over 3} {\texttt O}_{1,\Delta I=1/2}^{udsu}+{1 \over 3}{\texttt O}_{1,\Delta I=3/2}^{udsu}\;, &
{\texttt O}_3^{ddsd}&= {1 \over 3}{\texttt O}_{3,\Delta I=1/2}^{ddds}-{1 \over 3}{\texttt O}_{3,\Delta I=3/2}^{ddds}\;,
\end{align}
with their components of definite $\Delta I$ being given by
\begin{align}\nonumber
 {\texttt O}_{1,\Delta I=1/2}^{udsu} & = \Big[ \big( \overline{u_L}\gamma_\mu u_L\big) \big(\overline{s_R}\gamma^\mu d_R\big) - 2\big( \overline{u_L}\gamma_\mu d_L\big) \big(\overline{s_R}\gamma^\mu u_R\big) - \big( \overline{d_L}\gamma_\mu d_L\big) \big( \overline{s_R}\gamma^\mu d_R) \Big]J \,,
\\\nonumber
 {\texttt O}_{1,\Delta I=3/2}^{udsu} & = \Big[ \big( \overline{u_L}\gamma_\mu u_L\big) \big(\overline{s_R}\gamma^\mu d_R\big) + \big( \overline{u_L}\gamma_\mu d_L\big) \big(\overline{s_R}\gamma^\mu u_R\big) - \big( \overline{d_L}\gamma_\mu d_L\big) \big( \overline{s_R}\gamma^\mu d_R) \Big]J \,,
\\\nonumber
{\texttt O}_{3,\Delta I=1/2}^{ddsd} & = \Big[ \big(\overline{u_R}u_L\big) \big(\overline{s_R}d_L)+\big(\overline{u_R}d_L\big) \big(\overline{s_R}u_L)+2(\overline{d_R}d_L\big) \big(\overline{s_R}d_L\big) \Big]J \,,
\\
{\texttt O}_{3,\Delta I=3/2}^{ddsd} & = \Big[ \big(\overline{u_R}u_L\big) \big(\overline{s_R}d_L\big)+\big(\overline{u_R}d_L\big) \big(\overline{s_R}u_L)-(\overline{d_R}d_L\big) \big(\overline{s_R}d_L) \Big]J \,,
\end{align}
and similarly $\tilde{\texttt O}{}_1^{udsu}$ and $\tilde{\texttt O}{}_3^{ddsd}$.
\end{itemize}

\section{RG running of dim-6 four-quark operators for \boldmath$K\to\pi S(S)$\label{rge}}

The 1-loop QCD RG equations of the Wilson coefficients of the LEFT dim-6 quark operators relevant to the $K\to\pi S(S)$ transitions are given by~\cite{Jenkins:2017dyc}
\begin{align}
\mu{d\over d\mu}
\begin{pmatrix}
{\tt C}^{V,LL}_{ddsd} \smallskip \\ {\tt C}^{V1,LL}_{uusd} \smallskip \\ {\tt C}^{V8,LL}_{uusd} \smallskip \\ {\tt C}^{V1,LR}_{sduu} \smallskip \\ {\tt C}^{V8,LR}_{sduu} \smallskip \\
{\tt C}^{V1,LR}_{sddd} \smallskip \\ {\tt C}^{V8,LR}_{sddd}
\end{pmatrix}
=& -\frac{\alpha_s}{2\pi}
\begin{pmatrix}
-{20\over 9}& 0& -{1\over 18}& 0 &-{1\over 18}  & 0 & -{1\over 18} \smallskip \\
0& 0& -{4\over 3}   & 0 & 0  & 0 & 0 \smallskip \\
-{4\over 3}& -6& {5\over 3}  & 0 & -\frac{1}{3} &  0 & -\frac{1}{3} \smallskip \\
0& 0& 0& 0 &{4\over 3}  & 0 & 0 \smallskip \\
-{4\over 3}& 0& -\frac{1}{3}  & 6 &  -{22\over 3} &  0 & -\frac{1}{3} \smallskip \\
0& 0& 0& 0 & 0  &  0 &{4\over 3} \smallskip \\
-{4\over 3}& 0& -\frac{1}{3}& 0 & -\frac{1}{3}  & 6 & -{22\over 3}
\end{pmatrix}
\begin{pmatrix}
{\tt C}^{V,LL}_{ddsd} \smallskip \\ {\tt C}^{V1,LL}_{uusd} \smallskip \\ {\tt C}^{V8,LL}_{uusd} \smallskip \\ {\tt C}^{V1,LR}_{sduu} \smallskip \\ {\tt C}^{V8,LR}_{sduu} \smallskip \\
{\tt C}^{V1,LR}_{sddd} \smallskip \\ {\tt C}^{V8,LR}_{sddd}
\end{pmatrix} ,
\\
\mu{d\over d\mu}
\begin{pmatrix}
{\tt C}^{S1,LL}_{uusd} \smallskip \\
{\tt C}^{S8,LL}_{uusd} \smallskip \\
{\tt C}^{S1,LL}_{udsu} \smallskip \\
{\tt C}^{S8,LL}_{udsu}
\end{pmatrix}
=&
-\frac{\alpha_s}{2\pi}
\begin{pmatrix}
8 &  -{8\over 9}  &-{32\over 9}& -{56\over 27} \smallskip \\
-4 & -{8\over 3}& {8\over 3}&  -{22\over 9} \smallskip \\
-{32\over 9} & - {56\over 27} & 8 & -{8\over 9} \smallskip \\
{8\over 3} &-{22\over 9} & -4 & -{8\over 3}
\end{pmatrix}^{\vphantom{|_|^|}}
\begin{pmatrix}
{\tt C}^{S1,LL}_{uusd} \smallskip \\
{\tt C}^{S8,LL}_{uusd} \smallskip \\
{\tt C}^{S1,LL}_{udsu} \smallskip \\
{\tt C}^{S8,LL}_{udsu}
\end{pmatrix} .
\end{align}
The solutions to these equations between the electroweak scale, which we take to be $\mu=m_W^{}$, and the chiral symmetry breaking $\mu=\Lambda_\chi=4\pi F_\pi\simeq1.2\;$GeV are
\begin{align}
\begin{pmatrix}
{\tt C}^{V,LL}_{ddsd} \\ {\tt C}^{V1,LL}_{uusd}  \\ {\tt C}^{V8,LL}_{uusd}\\
{\tt C}^{V1,LR}_{sduu} \\ {\tt C}^{V8,LR}_{sduu} \\
{\tt C}^{V1,LR}_{sddd} \\ {\tt C}^{V8,LR}_{sddd}
\end{pmatrix}_{\!\!\mu=\Lambda_{\chi}}
=&
\begin{pmatrix}
 0.76 & 0.00 & -0.01 & -0.00 & -0.00 & -0.00 & -0.00 \\
 0.01 & 1.07 & -0.19 & 0.00 & 0.00 & 0.00 & 0.00 \\
 -0.16 & -0.86 & 1.31 & -0.01 & -0.03 & -0.01 & -0.03 \\
 -0.01 & 0.00 & -0.00 & 1.05 & 0.11 & -0.00 & -0.00 \\
 -0.09 & 0.01 & -0.03 & 0.51 & 0.43 & -0.01 & -0.02 \\
 -0.01 & 0.00 & -0.00 & -0.00 & -0.00 & 1.05 & 0.11 \\
 -0.09 & 0.01 & -0.03 & -0.01 & -0.02 & 0.51 & 0.43 \\
\end{pmatrix}
\begin{pmatrix}
{\tt C}^{V,LL}_{ddsd} \smallskip \\ {\tt C}^{V1,LL}_{uusd} \smallskip \\
{\tt C}^{V8,LL}_{uusd} \smallskip \\
{\tt C}^{V1,LR}_{sduu} \smallskip \\ {\tt C}^{V8,LR}_{sduu} \smallskip \\
{\tt C}^{V1,LR}_{sddd} \smallskip \\ {\tt C}^{V8,LR}_{sddd}
\end{pmatrix}_{\!\!\mu=m_W^{}} \,,
\\
\begin{pmatrix}
{\tt C}^{S1,LL}_{uusd} \smallskip \\
{\tt C}^{S8,LL}_{uusd} \smallskip \\
{\tt C}^{S1,LL}_{udsu} \smallskip \\
{\tt C}^{S8,LL}_{udsu}
\end{pmatrix}_{\!\!\mu=\Lambda_{\chi}}
=&
\begin{pmatrix}
 2.97 & -0.03 & -1.17 & -0.36 \smallskip \\
 -1.01 & 0.71 & 0.84 & -0.16 \smallskip \\
 -1.17 & -0.36 & 2.97 & -0.03 \smallskip \\
 0.84 & -0.16 & -1.01 & 0.71 \smallskip \\
\end{pmatrix}
\begin{pmatrix}
{\tt C}^{S1,LL}_{uusd} \smallskip \\
{\tt C}^{S8,LL}_{uusd} \smallskip \\
{\tt C}^{S1,LL}_{udsu} \smallskip \\
{\tt C}^{S8,LL}_{udsu}
\end{pmatrix}_{\!\!\mu=m_W^{}} \,.
\end{align}
All of these formulas are also valid for the chirality-flipped counterparts of the operators.

\section{Chiral structure and hadronization of quark operators for \boldmath$K\to\pi S(S)$}
\label{Appendix:3}

Here we collect the SU$(3)_L\times{\rm SU}(3)_R$ irreducible representations of the dim-6 four-quark operators examined in section\,\,\ref{k2ps} and the corresponding mesonic operators decomposed into their $\Delta I=1/2,3/2$ components.
Adopting the normalization convention of ref.\,\,\cite{Cirigliano:2017ymo} for the chiral realization of each operator,\footnote{Particularly, 
$(\overline{s_L}\gamma_\mu d_L)(\overline{s_L}\gamma^\mu d_L) \Rightarrow 
\frac{5}{12} g_{27\times1}^{} F_0^4 \tilde L_{\mu23}^{} \tilde L^\mu_{23}$ 
and 
$(\overline{s_L}\gamma_\mu d_L)(\overline{s_R}\gamma^\mu d_R) \Rightarrow 
\frac{1}{4} g_{8\times8}^{} F_0^4 \Sigma_{23}^{} \Sigma^\dagger_{23}$
among the operators with purely left-handed quarks and quarks of mixed chirality, respectively.
} we have
\begin{align}\nonumber
\textsl{\texttt O}^{V,LL}_{ddsd}|_{27\times1} & \,=\, {1\over 5}\left[(4\overline{d_L}\gamma^\mu d_L-\overline{s_L}\gamma^\mu s_L)(\overline{s_L}\gamma_\mu d_L)-(\overline{u_L}\gamma^\mu d_L)(\overline{s_L}\gamma_\mu u_L)\right]-{1\over 5}(\overline{q_L}\gamma^\mu q_L)(\overline{s_L}\gamma_\mu d_L)
\\ & \,\Rightarrow\,
{1\over 12}g_{27\times1}F_0^4\left[4 \tilde L_{\mu22}\tilde L^\mu_{23}-\tilde L_{\mu33}\tilde L^\mu_{23}-\tilde L_{\mu21}\tilde L^\mu_{13}\right]
\,\supset\, {1\over18}g_{27\times1}^{}\left(2{\cal Q}_{1/2}^V-5{\cal Q}_{3/2}^V \right) ,
\end{align}
\begin{align} \nonumber
\textsl{\texttt O}^{V,LL}_{ddsd}|_{8\times1}^{} & \,=\, {1\over 5}\left[(\overline{u_L}\gamma^\mu d_L)(\overline{s_L}\gamma_\mu u_L)-(\overline{u_L}\gamma^\mu u_L)(\overline{s_L}\gamma_\mu d_L)\right]+{2\over 5}(\overline{q_L}\gamma^\mu q_L)(\overline{s_L}\gamma_\mu d_L)
\\ & \,\Rightarrow\,
{1\over 12}g_{8\times1}^{}F_0^4\left[\tilde L_{\mu21}\tilde L^\mu_{13}-\tilde L_{\mu11}\tilde L^\mu_{23}\right]
\,\supset\, {1\over6}^{}g_{8\times1}^{}{\cal Q}_{1/2}^V \,, & ~~~
\end{align}
\begin{align} \nonumber
\textsl{\texttt O}^{V1,LL}_{uusd}|_{27\times1} & \,=\, {1\over 5}\left[ 3(\overline{u_L}\gamma^\mu u_L)(\overline{s_L}\gamma_\mu d_L)+2(\overline{u_L}\gamma^\mu d_L)(\overline{s_L}\gamma_\mu u_L)\right]-{1\over 5}(\overline{q_L}\gamma^\mu q_L)(\overline{s_L}\gamma_\mu d_L)
\\ & \,\Rightarrow\,
{1\over 12}g_{8\times1}^{}F_0^4\left[3\tilde L_{\mu11}\tilde L^\mu_{23}+2\tilde L_{\mu21}\tilde L^\mu_{13}\right]
\,\supset\, {1\over18}g_{27\times1}^{}\left({\cal Q}_{1/2}^V+5{\cal Q}_{3/2}^V \right) , & ~~
\end{align}
\begin{align} \nonumber
\textsl{\texttt O}^{V1,LL}_{uusd}|_{8\times1}^{} & \,=\, {2\over 5}\left[(\overline{u_L}\gamma^\mu u_L)(\overline{s_L}\gamma_\mu d_L)-(\overline{u_L}\gamma^\mu d_L)(\overline{s_L}\gamma_\mu u_L)\right]+{1\over 5}(\overline{q_L}\gamma^\mu q_L)(\overline{s_L}\gamma_\mu d_L)
\\ & \,\Rightarrow\,
{1\over6}^{}g_{8\times1}^{}F_0^4\left[\tilde L_{\mu11}\tilde L^\mu_{23}-\tilde L_{\mu21}\tilde L^\mu_{13}\right]
\,\supset\, -\frac{1}{3}g_{8\times1}^{}{\cal Q}_{1/2}^V \,, & ~~~
\end{align}
\begin{align}
\textsl{\texttt O}^{V8,LL}_{uusd}|_{27\times1} & \,=\, \frac{1}{3}\textsl{\texttt O}^{V1,LL}_{uusd}|_{27\times1} \,, & ~~~~~~ & & ~~~
\end{align}
\begin{align} \nonumber
\textsl{\texttt O}^{V8,LL}_{uusd}|_{8\times1}^{} & \,=\, {11\over 30}\left[(\overline{u_L}\gamma^\mu d_L)(\overline{s_L}\gamma_\mu u_L)-(\overline{u_L}\gamma^\mu u_L)(\overline{s_L}\gamma_\mu d_L)\right]+{1\over 15}(\overline{q_L}\gamma^\mu q_L)(\overline{s_L}\gamma_\mu d_L)
\\ & \,\Rightarrow\,
{11\over 72}g_{8\times1}^{}F_0^4\left[\tilde L_{\mu21}\tilde L^\mu_{13}-\tilde L_{\mu11}\tilde L^\mu_{23}\right]
\,\supset\, {11\over 36}g_{8\times1}^{}{\cal Q}_{1/2}^V \,, & ~~
\end{align}
\begin{align} \nonumber
\textsl{\texttt O}^{V1,LR}_{sduu}|_{8\times8}^{} & \,=\, (\overline{s_L}\gamma_\mu d_L)\left[ (\overline{u_R}\gamma^\mu u_R)-\frac{1}{3}(\overline{q_R}\gamma^\mu q_R)\right]
\\ & \,\Rightarrow\,
\frac{F_0^4}{4}g_{8\times8}^{}\Sigma_{21} \Sigma^\dagger_{13}
\,\supset\, {1\over6}g_{8\times8}^{}\left(2{\cal Q}_{1/2}^S+{\cal Q}_{3/2}^S \right) , & ~~~~~ & & ~~~
\end{align}
\begin{align} \nonumber
\textsl{\texttt O}^{V8,LR}_{sduu}|_{8\times8}^{} & \,=\, -{1\over6}^{}(\overline{s_L}\gamma_\mu d_L)\left[ (\overline{u_R}\gamma^\mu u_R)-\frac{1}{3}(\overline{q_R}\gamma^\mu q_R)\right]+{1\over 2}(\overline{s_L}\gamma_\mu d_L]\left\{ [\overline{u_R}\gamma^\mu u_R)-\frac{1}{3}[\overline{q_R}\gamma^\mu q_R)\right\}
\\ & \,\Rightarrow\,
-{1\over6}^{}\frac{F_0^4}{4}\left( g_{8\times8}^{}-3\tilde g_{8\times8}^{}  \right)\Sigma_{21} \Sigma^\dagger_{13}
\,\supset\, {1\over36}\left( 3\tilde g_{8\times8}^{} - g_{8\times8}^{} \right)\left(2{\cal Q}_{1/2}^S+{\cal Q}_{3/2}^S \right) ,
\end{align}
\begin{align} \nonumber
\textsl{\texttt O}^{V1,LR}_{sddd}|_{8\times8}^{} & \,=\, (\overline{s_L}\gamma_\mu d_L)\left[ (\overline{d_R}\gamma^\mu d_R)-\frac{1}{3}(\overline{q_R}\gamma^\mu q_R)\right]
\\ & \,\Rightarrow\,
\frac{F_0^4}{4}g_{8\times8}^{}\Sigma_{22} \Sigma^\dagger_{23}
\,\supset\, -{1\over12}g_{8\times8}^{}\left({\cal Q}_{1/2}^S+2{\cal Q}_{3/2}^S \right) , & \hspace{9em}
\end{align}
\begin{align} \nonumber
\textsl{\texttt O}^{V8,LR}_{sddd}|_{8\times8}^{} & \,=\, -{1\over6}^{}(\overline{s_L}\gamma_\mu d_L)\left[ (\overline{d_R}\gamma^\mu d_R)-\frac{1}{3}(\overline{q_R}\gamma^\mu q_R)\right]+{1\over2}(\overline{s_L}\gamma_\mu d_L]\left\{ [\overline{d_R}\gamma^\mu d_R)-\frac{1}{3}[\overline{q_R}\gamma^\mu q_R)\right\}
\\ & \,\Rightarrow\,
-{1\over6}^{}\frac{F_0^4}{4}\left( g_{8\times8}^{}-3\tilde g_{8\times8}^{}\right)\Sigma_{22} \Sigma^\dagger_{23}
\,\supset\, {1\over72}\left( g_{8\times8}^{}-3\tilde g_{8\times8}^{}\right)\left({\cal Q}_{1/2}^S+2{\cal Q}_{3/2}^S \right) ,
\end{align}
\begin{align} \nonumber
\textsl{\texttt O}^{S1,LL}_{uusd}|_{6\times\bar6}^{} & \,=\,
{1\over 2}\left[(\overline{u_R}u_L)(\overline{s_R}d_L)+(\overline{u_R}d_L)(\overline{s_R}u_L)\right]
\\ & \,\Rightarrow\,
{1\over2}\frac{F_0^4}{4}g_{6\times\bar6}^{} \left(\Sigma_{23}\Sigma_{11}+\Sigma_{13}\Sigma_{21}\right)
\,\supset\, -{1\over 24}g_{6\times\bar6}^{}\left( 5{\cal Q}_{1/2}^S+4{\cal Q}_{3/2}^S\right) , & \hspace{4em}
\end{align}
\begin{align} \nonumber
\textsl{\texttt O}^{S1,LL}_{uusd}|_{\bar3\times3} & \,=\, {1\over 2}\left[(\overline{u_R}u_L)(\overline{s_R}d_L)-(\overline{u_R}d_L)(\overline{s_R}u_L)\right]
\\ & \,\Rightarrow\,
{1\over2}\frac{F_0^4}{4} g_{\bar3\times 3} \left(\Sigma_{23}\Sigma_{11}-\Sigma_{13}\Sigma_{21}\right) \,\supset\, \frac{1}{8} g_{\bar3\times 3}{\cal Q}_{1/2}^S \,, & \hspace{11em}
\end{align}
\begin{align} \nonumber
\textsl{\texttt O}^{S8,LL}_{uusd}|_{6\times\bar6}^{} & \,=\,
-{1\over 12}\left[(\overline{u_R}u_L)(\overline{s_R}d_L)+(\overline{u_R}d_L)(\overline{s_R}u_L)  \right]
+{1\over 4}\left\{(\overline{u_R}u_L][\overline{s_R}d_L)+(\overline{u_R}d_L][\overline{s_R}u_L)  \right\}
\\ & \,\Rightarrow\,
\frac{F_0^4}{84} \left(g_{6\times\bar6}^{}-3\tilde g_{6\times\bar6}^{}  \right)\left(\Sigma_{23}\Sigma_{11}+\Sigma_{13}\Sigma_{21}\right)
\,\supset\, {g_{6\times\bar6}^{}-3\tilde g_{6\times\bar6}^{}\over 144} \left( 5{\cal Q}_{1/2}^S+4{\cal Q}_{3/2}^S\right) ,
\end{align}
\begin{align} \nonumber
\textsl{\texttt O}^{S8,LL}_{uusd}|_{\bar3\times3} & \,=\, -{1\over 12}\left[(\overline{u_R}u_L)(\overline{s_R}d_L)-(\overline{u_R}d_L)(\overline{s_R}u_L)  \right]
+{1\over 4}\left\{(\overline{u_R}u_L][\overline{s_R}d_L)-(\overline{u_R}d_L][\overline{s_R}u_L)  \right\}
\\ & \,\Rightarrow\,
-\frac{F_0^4}{48}\left( g_{\bar3\times3}-3\tilde g_{\bar3\times 3} \right)\left(\Sigma_{23}\Sigma_{11}-\Sigma_{13}\Sigma_{21}\right)
\,\supset\, {3\tilde g_{\bar3\times3} - g_{\bar3\times3}\over 48} {\cal Q}_{1/2}^S \,, & ~
\end{align}
\begin{align} \nonumber
\textsl{\texttt O}^{S1,LL}_{udsu}|_{6\times\bar6}^{} & \,=\, \textsl{\texttt O}^{S1,LL}_{uusd}|_{6\times\bar6}^{} \,, \hspace{7em}
\textsl{\texttt O}^{S1,LL}_{udsu}|_{\bar3\times3} \,=\, -\textsl{\texttt O}^{S1,LL}_{uusd}|_{\bar3\times3} \,,
\\
\textsl{\texttt O}^{S8,LL}_{udsu}|_{6\times\bar6}^{} & \,=\, \textsl{\texttt O}^{S8,LL}_{uusd}|_{6\times\bar6}^{} \,, \hspace{7em}
\textsl{\texttt O}^{S8,LL}_{udsu}|_{\bar3\times3} \,=\, -\textsl{\texttt O}^{S8,LL}_{uusd}|_{\bar3\times3} \,, & \hspace{7em}
\end{align}
where $q^{\rm T}=(u,d,s)$ and $\tilde L^\mu_{ij}=(\Sigma\partial^\mu\Sigma^\dagger)_{ij}$. For the chirality-flipped counterparts of these operators, the irreducible components and chiral realizations can be obtained from the above results by making the exchanges $L\leftrightarrow R$ and $\Sigma\leftrightarrow \Sigma^\dagger$.
We observe that among these operators $\textsl{\texttt O}^{V,LL}_{ddsd}|_{8\times1}^{}$, $\textsl{\texttt O}^{V1,LL}_{uusd}|_{8\times1}^{}$, $\textsl{\texttt O}^{V8,LL}_{uusd}|_{8\times1}^{}$, $\textsl{\texttt O}^{S1,LL}_{uusd,udsu}|_{\bar3\times3}$, and $\textsl{\texttt O}^{S8,LL}_{uusd,udsu}|_{\bar3\times3}$ generate exclusively $\Delta I=1/2$ interactions.


\end{document}